\documentclass[reprint,amsmath,amssymb,aps,pra,longbibliography]{revtex4-2}

\usepackage{graphicx}
\usepackage{dcolumn}
\usepackage{bm}
\usepackage{physics}
\usepackage{siunitx}
\usepackage{hyperref}

\newcommand*{\reffig}[1]{Fig.~\ref{#1}}
\newcommand*{\reftab}[1]{Table~\ref{#1}}
\newcommand*{\refeq}[1]{Eq.~(\ref{#1})}
\newcommand*{\refsec}[1]{Sec.~\ref{#1}}

\usepackage[english]{babel}
\babeladjust{autoload.bcp47 = on}

\begin{document}

\title{Parallel Hadamard Test}

\author{Soichiro Imamura}
\email{imamura-soichiro524@g.ecc.u-tokyo.ac.jp}
 
\affiliation{
 Department of Physics, The University of Tokyo, Tokyo 113-0033, Japan
}

\author{Synge Todo}
\email{wistaria@phys.s.u-tokyo.ac.jp}
 
\affiliation{
 Department of Physics, The University of Tokyo, Tokyo 113-0033, Japan\\
 Institute for Physics of Intelligence, The University of Tokyo, Tokyo 113-0033, Japan\\
 Institute for Solid State Physics, The University of Tokyo, Kashiwa, 277-8581, Japan
}

\date{\today}

\begin{abstract}

The Hadamard test is a fundamental building block widely used in many quantum computing algorithms.
It estimates the real or imaginary part of $\ev**{U}{\psi}$, 
where $\ket{\psi}$ is a quantum state and $U$ is a unitary operator.
In many algorithms, however, many such quantities must be estimated,
leading to a large number of distinct circuit types, long computational times, and high financial costs.
In this work, we propose the parallel Hadamard test, 
which combines multiple Hadamard tests into a single circuit type. 
We demonstrate how the parallel Hadamard test applies to three structural classes of workloads: 
arbitrary sets of unitary operators, prefix-product arrays, and Gram-matrix elements.
For each class, we compare the cost of the parallel Hadamard test with that of the conventional one.
Our unified approach significantly reduces the number of distinct circuit types, 
and can lower both computational time and financial costs
in regimes where fixed per-circuit overheads dominate the total cost.
In Gram-matrix workloads, it can also reduce the total number of shots when typical off-diagonal overlaps are small.
\end{abstract}

\maketitle

\section{Introduction}\label{sec:level1:Introduction}

Quantum computation is expected to solve problems that are intractable for classical hardware, 
such as prime factoring~\cite{shor_algorithms_1994,shor_polynomialtime_1999}, 
searching unsorted databases~\cite{grover_fast_1996, grover_quantum_1997}, 
material simulation~\cite{lloyd_universal_1996,abrams_quantum_1999,aspuru-guzik_simulated_2005}, 
linear algebraic operations~\cite{harrow_quantum_2009}, 
and machine learning~\cite{wiebe_quantum_2012,rebentrost_quantum_2014,cao_quantum_2017,kerenidis_quantum_2020}. 
These computational tasks are considered extremely difficult for classical machines, 
but quantum algorithms offer a promising route to address them efficiently and effectively.
In many quantum algorithms, it is often necessary to estimate the expectation value $\ev{U}{\psi}$, 
where $U$ is a unitary operator and $\ket{\psi}$ is a quantum state. 
This type of estimation is a common and essential step in many quantum procedures.
The Hadamard test (HT) is commonly used for estimating this quantity. 

On noisy intermediate-scale quantum (NISQ) devices, for example,
HT is used to estimate gradients in variational quantum algorithms (VQAs)~\cite{cerezo_variational_2021}, 
making it a useful primitive for optimizing variational circuits on near-term quantum hardware.
HT-like circuits also appear in many algorithms for early fault-tolerant quantum computing (Early-FTQC)~\cite{katabarwa_early_2024}, 
including quantum phase estimation (QPE) algorithms that require only a single ancilla qubit~\cite{dobsicek_arbitrary_2007,svore_faster_2013,ding_even_2023,ni_lowdepth_2023}.
These applications illustrate the broad utility of HT-type circuit structures in quantum algorithm design.
In addition, HT is used in quantum kernel methods~\cite{liu_rigorous_2021, paine_quantum_2023}. 
Thus, from NISQ devices to fully error-corrected quantum systems, 
HT remains a key component in quantum applications such as machine learning and scientific computing.

In these algorithms, however, we need to execute HT for a vast number of different unitary operators, 
which increases the number of distinct circuit types.
This overhead results in higher compilation costs to generate executable circuits on the actual quantum device, 
longer queuing delays when accessing cloud-based quantum devices, 
and higher usage fees associated with task submission~\cite{wierichs_general_2022}. 
These practical challenges can significantly impact the efficiency and affordability of quantum computation, 
especially in cloud-based environments, and become a bottleneck for large-scale applications.

Previous approaches to reducing the number of distinct circuit types have been developed mainly in the context of gradient estimation for VQAs.
Li \textit{et al.} proposed a strategy that exploits the decomposition of observables into Pauli strings~\cite{li_efficient_2024}.
Using the fact that mutually commuting observables can be measured simultaneously~\cite{verteletskyi_measurement_2020}, together with a technique for swapping the generators of parameterized gates with the observables, their method reduces the number of required circuit types.
Heidari \textit{et al.} focused on the dynamical Lie algebra determined by the structure of a parameterized quantum circuit (PQC)~\cite{heidari_efficient_2024}.
By imposing conditions on the dynamical Lie algebra and on the Hilbert-Schmidt norms of the observables, they also demonstrated a reduction in the number of circuit types.
However, these methods are specialized to gradient estimation in VQAs and are fundamentally different from the approach proposed in this work.

In this work, we propose a novel method, the parallel Hadamard test (PHT), which enables numerous HTs to be executed in parallel using a single quantum circuit,
thereby reducing the number of distinct circuit types. 
The PHT framework enables the simultaneous evaluation of multiple expectation values
and is applicable across hardware regimes, including NISQ, Early-FTQC, and fault-tolerant quantum computing (FTQC). 
In the most straightforward implementation of PHT, the number of required ancilla qubits increases as the number of quantities to be estimated increases.
However, when the hardware supports mid-circuit measurement and reset (MCMR), 
this requirement can be reduced to only two or three ancilla qubits.
This reduced ancilla overhead makes the method more practical for implementation on existing devices.
Because PHT preserves the basic structure of the standard HT, 
it can be incorporated into a broad range of quantum algorithms without substantial modifications.

The remainder of this paper is organized as follows.
In \refsec{sec:level1:HT}, we first briefly summarize the standard HT and its representative applications.
In \refsec{sec:level1:PHT}, we detail the proposed method, PHT, and its circuit construction.
\refsec{sec:level1:CostComparison} provides a comparative analysis of PHT against HT.
Finally, \refsec{sec:level1:discussion_and_conclusion} concludes the paper with a discussion of limitations and future outlook.

\section{Hadamard Test}\label{sec:level1:HT}

\begin{figure}[bt]
\includegraphics[width = 0.2\textwidth]{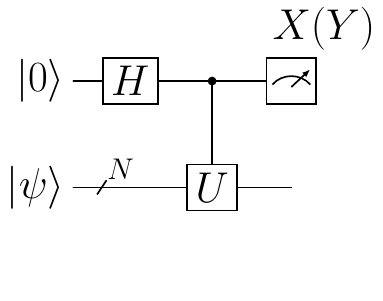}
\caption{\label{fig:HT_general} 
Quantum circuit for HT.
The ancilla qubit is measured in the Pauli $X$ ($Y$) basis to estimate the real (imaginary) part of $\ev**{U}{\psi}$.
}
\end{figure}

HT is a quantum algorithm for estimating either the real or imaginary part of $\ev{U}{\psi}$
using the circuit shown in \reffig{fig:HT_general}.
When the ancilla qubit is measured in the Pauli $X$ basis, 
its expectation value gives the real part of $\ev**{U}{\psi}$:
\begin{align}
  \ev*{X} =  \Re[\ev**{U}{\psi}]. \label{eq:HT_general}
\end{align}
Similarly, the imaginary part is obtained by measuring the ancilla qubit in the Pauli $Y$ basis.

As the measurement outcome of the ancilla qubit is either $+1$ or $-1$, the statistical error of HT is given by
\begin{align}
  \sigma = \sqrt{\frac{1 - x^2}{N_{\mathrm{shot}}}},
\end{align}
where $N_{\mathrm{shot}}$ is the number of shots and $x$ is the expected value of the quantity to be estimated.
In other words, the number of shots required to achieve a target precision $\varepsilon$ is determined by the relation
\begin{align}
    N_{\mathrm{shot}} = \frac{1 - x^2}{\varepsilon^2}. \label{eq:HT_shot}
\end{align}

In the following subsections, we present several representative applications of HT.

\subsection{Gradient estimation in variational quantum algorithms}\label{sec:level2:gradient_estimation}

VQAs are hybrid quantum-classical algorithms that use PQCs 
to obtain approximate solutions to optimization problems~\cite{cerezo_variational_2021}. 
A representative example is the variational quantum eigensolver (VQE)~\cite{peruzzo_variational_2014,tilly_variational_2022}, 
which seeks an approximate ground state of a given Hamiltonian $\mathcal{H}$ by minimizing its energy expectation value. 
This optimization proceeds iteratively: 
quantum states are prepared using PQCs, the cost function is evaluated on quantum hardware, and the circuit parameters are updated by a classical optimizer.

\begin{figure}[bt]
\includegraphics[width = 0.3\textwidth]{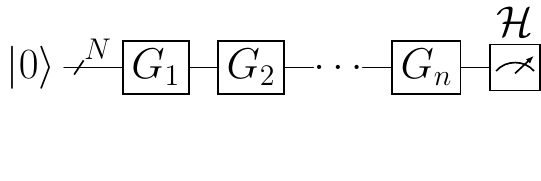}
\caption{\label{fig:PQC} 
Quantum circuit for VQE. 
Although the circuit is depicted as being measured with respect to the Hamiltonian $\mathcal{H}$,
in practice, $\mathcal{H}$ is decomposed into a sum of tensor products of Pauli operators, called Pauli strings. 
These Pauli strings are grouped into mutually commuting sets, 
and a separate quantum circuit is executed for each group~\protect\cite{verteletskyi_measurement_2020}.
}
\end{figure}

Although parameter updates can be performed without gradient information~\cite{nelder_simplex_1965,nakanishi_sequential_2020,ostaszewski_structure_2021,watanabe_optimizing_2023,wada_simulating_2022,parrish_jacobi_2019},
gradient-based optimization is generally expected to achieve faster convergence~\cite{harrow_lowdepth_2021}. 
Efficient gradient estimation is therefore crucial for practical VQE implementations. 

We consider an $N$-qubit PQC consisting of a sequence of $n$ unitary gates, as illustrated in \reffig{fig:PQC}.
We assume that each gate is parameterized by $\theta_j$ as
\begin{align}
  G_j(\theta_j) &= \exp\qty(-i\theta_j U_j/2) \quad \text{for} ~ j = 1,2,\dots , n, \label{eq:gate}
\end{align}
where the generator $U_j$ is Hermitian and unitary, i.e.,
\begin{align}
  U_j^\dagger &= U_j, ~ U_j^2 = I  \quad \text{for} ~ j = 1,2,\dots , n. \label{eq:generator}
\end{align}
Using the shorthand notation $G_{b:a}\equiv G_b G_{b-1}\cdots G_a$, 
the cost function, i.e., the energy expectation value of the Hamiltonian, and its gradient with respect to $\theta_j$ are given by
\begin{align}
  \ev{\mathcal{H}} &= \ev**{G_{n:1}^\dagger \mathcal{H} G_{n:1}}{0},\\
  \pdv{\ev{\mathcal{H}}}{\theta_j} &= \Im \qty[\ev**{G_{n:1}^\dagger \mathcal{H} G_{n:j+1} U_j G_{j:1}}{0}]\nonumber\\
  &= \Im\left[\qty(\bra{0}G_{n:1}^\dagger)\mathcal{H}\bigl(G_{n:j+1} U_j G_{n:j+1}^\dagger\bigr)\qty(G_{n:1}\ket{0})\right],\label{eq:gradient}
\end{align}
respectively.

\begin{figure}[bt]
\centering
\includegraphics[width = 0.4\textwidth]{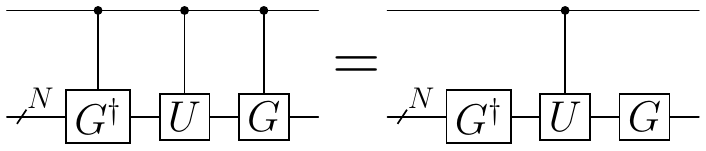}
\caption{Circuit identity for simplifying controlled blocks. 
When a controlled-$G$ gate and its inverse controlled-$G^\dagger$ gate with the same control qubit sandwich another controlled-$U$ gate, 
the controls on $G$ and $G^\dagger$ are redundant. 
Thus, replacing the controlled-$G$ and controlled-$G^\dagger$ gates with $G$ and $G^\dagger$ preserves the overall unitary transformation.}
\label{fig:level3:HTGradientCircuitIdentity}
\end{figure}

\begin{figure}[bt]
\includegraphics[width = 0.45\textwidth]{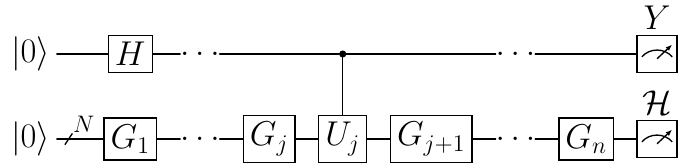}
\caption{\label{fig:VQE_HT} 
Quantum circuit for gradient estimation using HT.
The ancilla qubit is measured in the Pauli $Y$ basis because the gradient is given by the imaginary part of $\ev**{G_{n:1}^\dagger \mathcal{H} G_{n:j+1} U_j G_{j:1}}{0}$.
}
\end{figure}

HT provides a direct method for gradient estimation~\cite{guerreschi_practical_2017,romero_strategies_2018,schuld_evaluating_2019,li_efficient_2024}.
Although the system qubits are not measured in \reffig{fig:HT_general}, 
we can include a measurement of the Hamiltonian $\mathcal{H}$ on the system register.
When the ancilla qubit is measured in the Pauli $Y$ basis, 
the corresponding expectation value is
\begin{align}
  \ev{Y \otimes \mathcal{H}} = \Im[\ev{\mathcal{H}U}{\psi}].\label{eq:vqe_ht_im}
\end{align}
By setting $\ket{\psi}:=G_{n:1}\ket{0}$ and $U:=G_{n:j+1} U_j G_{n:j+1}^\dagger$ in \reffig{fig:HT_general},
and by applying the circuit identity shown in \reffig{fig:level3:HTGradientCircuitIdentity},
the gradient in \refeq{eq:gradient} can be estimated using the HT circuit shown in \reffig{fig:VQE_HT}.

An alternative approach to gradient estimation is the parameter-shift rule (PSR)~\cite{mitarai_quantum_2018,schuld_evaluating_2019,hubregtsen_singlecomponent_2022,crooks_gradients_2019,izmaylov_analytic_2021,wierichs_general_2022},
but both HT and PSR require distinct quantum circuits for different parameters. 
As the system size or circuit complexity increases, the number of parameters grows, 
leading to a corresponding increase in the number of required circuit types.

\subsection{Quantum phase estimation for Early-FTQC}\label{sec:level2:qpe_on_early_ftqc}

\begin{figure}[bt]
\includegraphics[width = 0.45\textwidth]{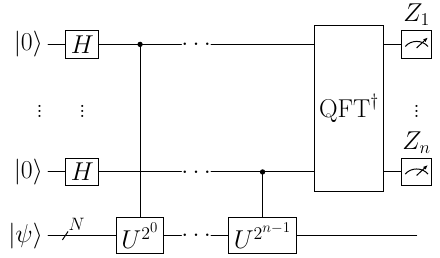}
\caption{\label{fig:QPE_naive} 
Quantum circuit for standard QPE~\protect\cite{nielsen_quantum_2010}. 
QFT$^\dagger$ denotes the inverse quantum Fourier transform.
The input state $\ket{\psi}$ is an eigenstate of the unitary operator $U$, satisfying $U \ket{\psi} = e^{2\pi i\varphi} \ket{\psi}$.
}
\end{figure}

QPE is one of the most important quantum algorithms 
for achieving quantum advantage across a range of applications.
It enables precise eigenphase estimation of a unitary operator, 
and when the operator represents Hamiltonian dynamics, the eigenphase encodes an energy eigenvalue.
However, standard QPE~\cite{nielsen_quantum_2010} 
requires a large number of ancilla qubits and a deep circuit, as shown in \reffig{fig:QPE_naive}.
NISQ devices are too noisy to execute such circuits, while FTQC remains out of reach.
Recently, there has been significant interest in Early-FTQC~\cite{katabarwa_early_2024},
a regime that lies between NISQ computing and FTQC.
Early-FTQC still has limitations in terms of qubit count and circuit depth,
but it can handle more qubits and deeper circuits than NISQ devices 
through the use of error mitigation and partial quantum error correction techniques.
This has motivated many studies on QPE algorithms that require only a single ancilla qubit, 
making them more suitable for Early-FTQC systems~\cite{dobsicek_arbitrary_2007,svore_faster_2013,ding_even_2023,ni_lowdepth_2023,kshirsagar_proving_2024,lin_heisenberglimited_2022,wan_randomized_2022,dutkiewicz_heisenberglimited_2022,obrien_quantum_2019}.
Extensive research has also focused on comparing and evaluating these algorithms~\cite{toshio_practical_2025,nelson_assessment_2024,liang_modeling_2024,ku_benchmarking_2025}.

\begin{figure}[bt]
\includegraphics[width = 0.2\textwidth]{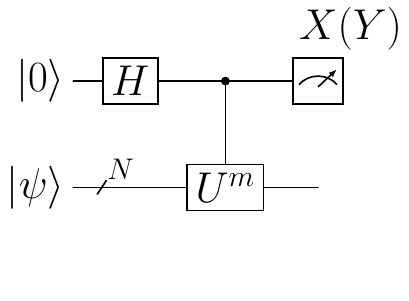}
\caption{\label{fig:QCELS}
Quantum circuit for QCELS. 
The ancilla qubit is measured in the Pauli $X$ ($Y$) basis to estimate the real (imaginary) part of $\ev**{U^m}{\psi}$.
}
\end{figure}

As a representative QPE algorithm tailored to Early-FTQC platforms, we consider
the quantum complex exponential least squares (QCELS)~\cite{ding_even_2023}, 
whose potential for quantum advantage has been investigated in Ref.~\cite{toshio_practical_2025}.
This algorithm uses an HT quantum circuit as shown in \reffig{fig:QCELS}.
Here, $U$ is the unitary operator whose eigenphase we want to estimate
and $\ket{\psi}$ is an input quantum state that has sufficient overlap with an eigenstate of $U$.
First, we estimate $\ev**{U^m}{\psi}$ by HT for each $m = 1, \dots, M$, and denote the estimate by $\hat{z}_m$. 
Next, we define a loss function as follows:
\begin{align}
  L(r, \varphi) = \frac{1}{M} \sum_{m=1}^{M} \abs{\hat{z}_m - r e^{2\pi i m \varphi}}^2. \label{eq:QCELS_loss}
\end{align}
When the number of shots for each circuit is sufficiently large, one of the global minima of the loss function corresponds to the eigenphase.
By classically optimizing the parameters $r$ and $\varphi$, we can estimate the eigenphase.
Thus, QCELS enables QPE on Early-FTQC platforms.

To achieve higher accuracy in estimating the eigenphase, 
one must modify the powers or evolution times of $U$ and execute another set of HT circuits~\cite{ding_even_2023}. 
To estimate the eigenphase with target precision $\varepsilon$, this process must be repeated $\lceil \log_2(1/\varepsilon) \rceil + 1$ times. 
Thus, QCELS requires the use of many distinct circuit types for HT.

\subsection{Quantum kernel method}\label{sec:level2:quantum_kernel_method}

\begin{figure}[bt]
\includegraphics[width = 0.3\textwidth]{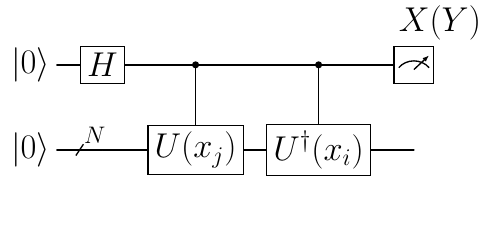}
\caption{\label{fig:kernel} 
Quantum circuit for evaluating a quantum kernel. 
The ancilla qubit is measured in the Pauli $X$ ($Y$) basis to estimate the real (imaginary) part of $\ev{U^\dagger(x_i)U(x_j)}{0^{\otimes N}}$.
}
\end{figure}

In supervised learning, feature maps are used to reveal patterns that are not easily distinguishable in the original data space. 
By mapping inputs to a higher-dimensional space, they can make classification tasks easier to solve with linear decision boundaries. 
This idea forms the basis of support vector machines (SVMs)~\cite{vapnik_nature_2000}. 

Consider a general feature map $\Phi: \mathcal{X} \rightarrow \mathcal{F}$, 
where $\mathcal{F}$ is a feature space equipped with an inner product $\langle \cdot, \cdot \rangle_{\mathcal{F}}$. 
To construct a linear classifier in $\mathcal{F}$, we solve a convex quadratic optimization problem involving 
slack variables $\xi_i \geq 0$, a regularization parameter $\lambda > 0$, and a training dataset 
$\{(x_i, y_i)\}_{i=1}^M \subseteq \mathbb{R}^d \times \{\pm 1\}$: 
\begin{align}
  &\min_{w, \xi} \qty(\frac{1}{2} \|w\|_{\mathcal{F}}^2 + \frac{\lambda}{2} \sum_{i=1}^M \xi_i^2) \nonumber\\ 
  &\text{s.t.} \quad y_i \langle \Phi(x_i), w \rangle_{\mathcal{F}} \geq 1 - \xi_i \quad \text{for $i = 1, \ldots, M$}.
\end{align}
The goal is to find a hyperplane $w$ in $\mathcal{F}$ 
such that the classifier $y = \text{sign}(\langle \Phi(x), w \rangle_{\mathcal{F}})$ separates the data with a large margin, 
while allowing some misclassification controlled by the slack variables.

The challenge arises when $\mathcal{F}$ is very high-dimensional, 
where direct computation of the optimal hyperplane can become infeasible. 
SVMs address this issue through the dual formulation of the optimization problem, 
which is formulated in terms of the kernel function $K(x_i, x_j) = \langle \Phi(x_i), \Phi(x_j) \rangle_{\mathcal{F}}$: 
\begin{align}
  \max_{\alpha \geq 0} \Biggl( \sum_{i=1}^M \alpha_i - \frac{1}{2} \sum_{i,j=1}^M \alpha_i \alpha_j y_i y_j K(x_i, x_j) - \frac{1}{2\lambda} \sum_{i=1}^{M} \alpha_i^2 \Biggr).
\end{align}
This dual formulation yields a classifier of the form $y = \text{sign}\left(\sum_{i=1}^M \alpha_i y_i K(x, x_i)\right)$, 
and, by strong duality, is mathematically equivalent to the original problem. 
Crucially, it enables computations in high-dimensional feature spaces, provided that the kernel function can be evaluated efficiently.

Quantum kernel methods replace classical kernel functions with estimates computed on a quantum computer~\cite{liu_rigorous_2021, paine_quantum_2023}. 
Compared with many classical feature maps, quantum feature maps can provide highly expressive feature spaces, 
which may help identify complex patterns in data~\cite{havlicek_supervised_2019}.
Specifically, the kernel function is defined as
\begin{align}
  K(x_i, x_j) 
  &= \abs{\braket{\psi(x_i)}{\psi(x_j)}}^2 \nonumber\\
  &= \abs{\ev{U^\dagger(x_i)U(x_j)}{0^{\otimes N}}}^2, \label{eq:quantum_kernel}
\end{align}
where the quantum hardware is used solely to estimate the kernel values, 
while the subsequent optimization is carried out on a classical computer. 
This hybrid approach, known as a quantum kernel method, has demonstrated speedups for the discrete logarithm problem~\cite{liu_rigorous_2021}. 
Moreover, its applicability extends beyond classification to regression problems and differential equations~\cite{paine_quantum_2023}.

One approach to estimating the overlaps needed for quantum kernel values is to use HT circuits, as shown in \reffig{fig:kernel}.
For a dataset of $M$ data points, however, the number of off-diagonal kernel elements that must be estimated grows as $M(M-1)/2$. 
Thus, as the dataset size increases, this quadratic scaling requires many distinct quantum circuits to be constructed and executed.

\section{Parallel Hadamard Test}\label{sec:level1:PHT}

To reduce the overhead caused by the proliferation of distinct circuit types,
we propose a method for parallelizing multiple HTs within a single quantum circuit.
We call this method the parallel Hadamard test (PHT).

\begin{figure}[bt]
\includegraphics[width = 0.4\textwidth]{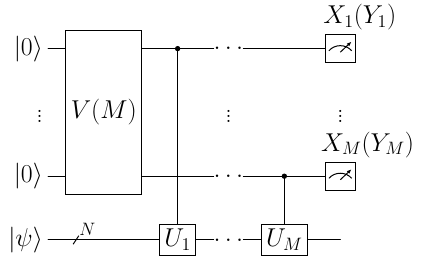}
\caption{\label{fig:PHT_general} 
Basic circuit structure of PHT.
The design of $V(M)$ depends on the quantities to be estimated.
Controlled unitary gates are inserted by shifting the control qubit one by one.
The ancilla qubits are measured in the Pauli $X$ or $Y$ basis.
}
\end{figure}

The basic circuit structure of PHT is shown in \reffig{fig:PHT_general}. 
The key idea is to prepare an appropriate ancilla state using the $V$ gate. 
The computational-basis components of this ancilla state determine which controlled unitary gates act on each branch of the superposition.
Thus, by designing the ancilla state appropriately, PHT selects the required unitary operators, or ordered products of them, within a single circuit.
The design of $V$ therefore depends on the form of the quantities to be estimated, 
and concrete examples are presented in the following sections.

In the following, we classify the design of the $V$ gate into three representative patterns, each corresponding to a different class of applications:
\begin{itemize}
    \item \textbf{Arbitrary set of unitary operators:} The most general setting, in which the goal is to estimate quantities of the form $\ev**{U_m}{\psi}$ 
    for distinct unitary operators $U_m$ $(m=1,2,\ldots,M)$. Gradient estimation in variational quantum algorithms such as VQE falls into this category.
    \item \textbf{Prefix-product array:} A setting in which the goal is to estimate expectation values of sequential products of unitary operators, 
    e.g., $\ev**{U_m U_{m-1} \cdots U_1}{\psi}$ $(m=1,2,\ldots,M)$. QCELS is a representative example with this structure.
    \item \textbf{Gram-matrix elements:} A setting in which pairs of unitary operators are selected from a given set to evaluate overlaps of the form $\ev**{U_i^\dagger U_j}{\psi}$ $(1\le i<j\le M)$. 
    Quantum kernel methods typically require this type of computation.
\end{itemize}
These three patterns illustrate how PHT can be adapted to different algorithmic structures through the design of the $V$ gate.

\subsection{Arbitrary set of unitary operators}\label{sec:level2:PHT_arbitrary_set_of_unitary_operators}

First, we consider the most general setting, 
in which the goal is to estimate $M$ expectation values of the form $\ev**{U_m}{\psi}$ for $m = 1, 2, \dots, M$. 
For this purpose, we prepare the $M$-qubit ancilla state $\ket{+W}$, defined as
\begin{align}
  &\ket{+W} \equiv \alpha  \ket{0}^{\otimes M} + \beta \ket{W},
\end{align}
where
\begin{align}
  \alpha, \beta &\in \mathbb{C}, \quad \abs{\alpha}^2 + \abs{\beta}^2 = 1,\label{eq:plus_W_state_coeff}\\
  \ket{W} &= \frac{1}{\sqrt{M}}\qty(\ket{100\dots0} + \ket{010\dots 0} + \cdots + \ket{000\dots 1})\nonumber\\
          &= \frac{1}{\sqrt{M}}\sum_{m=1}^{M} \ket{e_m}_M.\label{eq:W_state}
\end{align}
Here, $\ket{W}$ is called the W state~\cite{dur_three_2000}, and $\ket{e_m}_M$ is the one-hot basis state defined by
\begin{align}
  \ket{e_m}_M &\equiv \ket{0}^{\otimes (m-1)} \otimes \ket{1} \otimes \ket{0}^{\otimes (M-m)}.
\end{align}
We denote by $V_{+W}$ the gate that prepares $\ket{+W}$ from the all-zero state:
\begin{align}
  V_{+W}(M)\ket{0}^{\otimes M} = \ket{+W}.
\end{align}

The state $\ket{+W}$ contains no computational-basis states with more than one excitation among the $M$ ancilla qubits.
This one-hot structure ensures that, on each branch of the superposition, at most one unitary operator is selected.
After applying the $V_{+W}(M)$ gate, the full quantum state becomes
\begin{align}
    &\qty(\alpha \ket{0}^{\otimes M} + \beta \ket{W}) \otimes \ket{\psi}\nonumber\\
  = &\alpha \ket{0}^{\otimes M} \otimes \ket{\psi} + \frac{\beta}{\sqrt{M}} \sum_{m=1}^{M} \ket{e_m}_M \otimes \ket{\psi}.
\end{align}
After the controlled unitary gates are applied, the full quantum state $\ket{\Psi}$ becomes
\begin{align}
    \ket{\Psi} = \alpha\ket{0}^{\otimes M} \otimes \ket{\psi} + \frac{\beta}{\sqrt{M}} \sum_{m=1}^{M} \ket{e_m}_M \otimes U_m \ket{\psi}.
\end{align}
When the $m$-th ancilla qubit is measured in the Pauli $X$ basis,
its expectation value is
\begin{align}
  \ev**{X_m}{\Psi} &= \frac{2}{\sqrt{M}}\Re \qty[\alpha^* \beta \ev**{U_m}{\psi}]\nonumber\\
  &\text{for $m = 1, 2, \dots, M$}.
\end{align}
Thus, the real part of $\ev**{U_m}{\psi}$ is obtained as
\begin{align}
  \Re \qty[\ev**{U_m}{\psi}] = \frac{1}{\alpha^* \beta} \frac{\sqrt{M}}{2} \ev*{X_m} \quad \text{for $m = 1, 2, \dots, M$}
\end{align}
provided that $\alpha^*\beta$ is a nonzero real number.
Similarly, when the $m$-th ancilla qubit is measured in the Pauli $Y$ basis,
its expectation value is
\begin{align}
  \ev**{Y_m}{\Psi} &= \frac{2}{\sqrt{M}}\Im \qty[\alpha^* \beta \ev**{U_m}{\psi}]\nonumber\\
  &\text{for $m = 1, 2, \dots, M$}.
\end{align}
Under the same condition on $\alpha^*\beta$, the imaginary part is obtained as
\begin{align}
  \Im \qty[\ev**{U_m}{\psi}] = \frac{1}{\alpha^* \beta} \frac{\sqrt{M}}{2} \ev*{Y_m} \quad \text{for $m = 1, 2, \dots, M$}.
\end{align}
For $M=1$ and $\alpha=\beta=1/\sqrt{2}$, PHT reduces to the standard HT.

\begin{figure*}[bt]
\includegraphics[width = 0.75\textwidth]{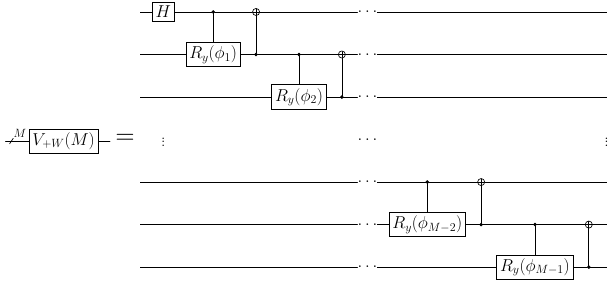}
\caption{\label{fig:V_plusW} 
Quantum circuit implementing the $V_{+W}(M)$ gate.
Here, $\phi_i = 2\arccos\qty(\sqrt{\frac{1}{M - i + 1}})$.
The initial Hadamard gate corresponds to $\alpha=\beta=1/\sqrt{2}$; for general $\alpha$ and $\beta$, use a suitable single-qubit rotation.
}
\end{figure*}

One implementation of the $V_{+W}$ gate uses controlled $R_y$ gates and CNOT gates arranged as shown in \reffig{fig:V_plusW}.
The rotation angles for the controlled $R_y$ gates are defined as
\begin{align}
  \phi_i = 2\arccos\qty(\sqrt{\frac{1}{M - i + 1}}) \quad \text{for $i = 1, 2, \ldots, M-1$}.
\end{align}
This sequential construction is compatible with MCMR, as discussed in \refsec{sec:level2:mid_circuit},
and is suitable for hardware platforms with limited qubit connectivity, such as superconducting quantum devices, 
where two-qubit gates are typically restricted to neighboring qubits.

\begin{figure}[bt]
\includegraphics[width = 0.45\textwidth]{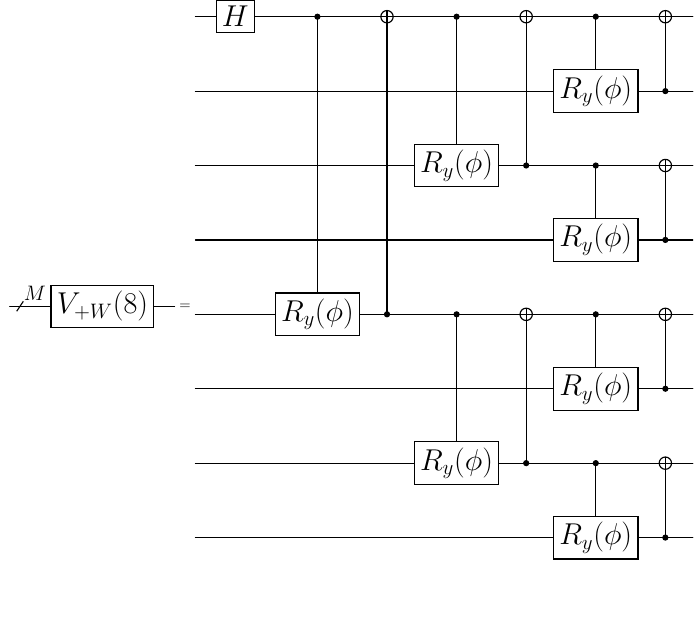}
\caption{\label{fig:V_gate_tree} 
Tree-like quantum circuit implementing the $V_{+W}(M)$ gate for $M=8$ qubits.
Here, $\phi = \pi/2$.
The initial Hadamard gate corresponds to $\alpha=\beta=1/\sqrt{2}$; for general $\alpha$ and $\beta$, use a suitable single-qubit rotation.
}
\end{figure}

An alternative implementation of the $V_{+W}$ gate uses a tree-like structure~\cite{cruz_efficient_2019}, as shown in \reffig{fig:V_gate_tree}.
Its circuit depth scales logarithmically with the number of qubits, 
and it can be constructed even when the number of qubits is not a power of two~\cite{cruz_efficient_2019}.
This method is more depth-efficient than the linear nearest-neighbor construction when the required connectivity is available.
However, the MCMR-based ancilla reduction discussed below relies on the sequential construction in \reffig{fig:V_plusW}, rather than this logarithmic-depth tree-like construction.

The estimators above give the correct expectation values as long as $\alpha^*\beta$ is a nonzero real number.
However, their statistical errors depend on the choice of $\alpha$ and $\beta$.
Measurement outcomes in the Pauli $X_m$ basis are either $-1$ or $1$.
For one shot, the statistical error $\sigma$ of the estimator for
$x=\Re \qty[\ev**{U_m}{\psi}]$ is
\begin{align}
  \sigma = \sqrt{\frac{1}{(\alpha^* \beta)^2} \frac{M}{4} - x^2}.
\end{align}
Thus, minimizing the statistical error is equivalent to maximizing $|\alpha^*\beta|$.
Under the above condition, we can take $\alpha$ and $\beta$ to be real and positive without loss of generality.
Using \refeq{eq:plus_W_state_coeff} and the arithmetic-geometric mean inequality, we obtain
\begin{align}
  \alpha^* \beta = \sqrt{\alpha^2 \beta^2} \leq \frac{\alpha^2 + \beta^2}{2} = \frac{1}{2}.
\end{align}
Thus, the maximum value of $\alpha^* \beta$ is $1/2$, achieved by setting $\alpha = \beta = 1/\sqrt{2}$.
This value is used throughout the remainder of this paper.
With this choice, the $V_{+W}(M)$ gate can be constructed as shown in \reffig{fig:V_plusW}.
The statistical error when we perform $N_{\mathrm{shot}}$ shots is given by
\begin{align}
  \sigma = \sqrt{\frac{M - x^2}{N_{\mathrm{shot}}}}. \label{eq:arbitrary_statistical_error}
\end{align}
Then, to achieve the target precision parameter $\varepsilon$, 
we need 
\begin{align}
    N_{\mathrm{shot}} = \frac{M - x^2}{\varepsilon^2} \label{eq:arbitrary_number_of_shots}
\end{align}
shots.
This reflects the increased single-estimator variance of PHT compared with the standard HT in \refeq{eq:HT_shot}.
In exchange, the same PHT circuit produces all $M$ estimates, reducing the number of distinct circuit types by a factor of $M$.

\begin{figure*}[bt]
\centering
\includegraphics[width = 0.95\textwidth]{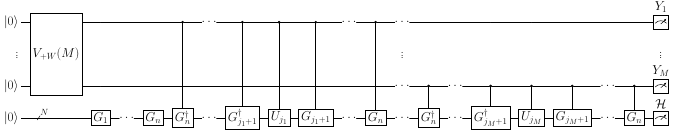}
\caption{\label{fig:VQE_PHT_naive} Naive PHT circuit for gradient estimation in VQE.}
\end{figure*}

\begin{figure*}[bt]
\centering
\includegraphics[width = 0.85\textwidth]{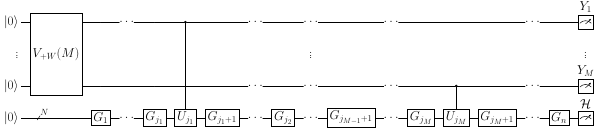}
\caption{\label{fig:VQE_PHT_optimized} PHT circuit for gradient estimation in VQE using the circuit identity in \reffig{fig:level3:HTGradientCircuitIdentity}.}
\end{figure*}

Regarding circuit depth, PHT is not automatically advantageous over a straightforward sequential arrangement of the original HT circuits. 
However, in some applications, the PHT circuit can be made substantially shallower. 
As a concrete example, we consider VQE gradient estimation, introduced in \refsec{sec:level2:gradient_estimation}.
A naive application of PHT gives the circuit shown in \reffig{fig:VQE_PHT_naive}.
Here, we order the target indices as $1\le j_1<j_2<\cdots<j_M\le n$.
In addition to the gates in the original PQC ($G_{n:1}$),
this construction requires many controlled versions of the PQC gates. 
By contrast, exploiting the circuit identity in \reffig{fig:level3:HTGradientCircuitIdentity} significantly reduces the circuit depth, as illustrated in \reffig{fig:VQE_PHT_optimized}. 
In the optimized PHT circuit, each gate in the original PQC is applied only once while all target quantities are estimated in parallel.
In contrast, a straightforward concatenation of the per-term HT circuits in \reffig{fig:VQE_HT} requires applying the entire PQC for each term. 
Thus, the PHT-based construction can yield a substantially shallower circuit.

\begin{figure*}[bt]
\centering
\includegraphics[width = 0.7\textwidth]{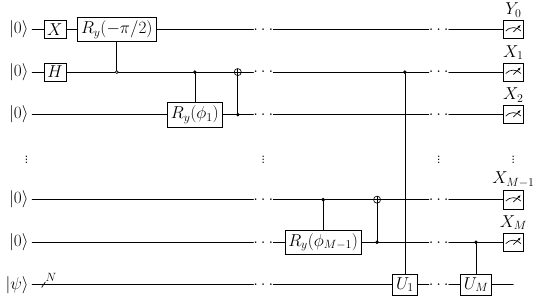}
\caption{
PHT circuit for simultaneously estimating the real and imaginary parts of $\ev**{U_m}{\psi}$ for $m = 1, 2, \dots, M$.
Here, $\phi_i = 2\arccos\qty(\sqrt{\frac{1}{M - i + 1}})$, and the open circle denotes a negative control.
}
\label{fig:arbitrary_simultaneously}
\end{figure*}

Moreover, PHT can estimate both the real and imaginary parts of $\ev**{U_m}{\psi}$ for all $m = 1, 2, \dots, M$ simultaneously with a single circuit.
The circuit shown in \reffig{fig:arbitrary_simultaneously} uses one additional ancilla qubit.
The real and imaginary parts are obtained from the corresponding ancilla observables as
\begin{align}
  \Re \qty[\ev**{U_m}{\psi}] &= \sqrt{2M} \ev{X_m}, \label{eq:arbitrary_simultaneously_real} \\
  \Im \qty[\ev**{U_m}{\psi}] &= \sqrt{2M} \ev{Y_0 X_m} \label{eq:arbitrary_simultaneously_imaginary} \\
  &\text{for $m = 1, 2, \dots, M$}. \nonumber
\end{align}
The statistical error after $N_{\mathrm{shot}}$ shots is
\begin{align}
  \sigma = \sqrt{\frac{2M - x^2}{N_{\mathrm{shot}}}}. \label{eq:arbitrary_simultaneously_statistical_error}
\end{align}
Thus, to achieve target precision $\varepsilon$, 
we need
\begin{align}
    N_{\mathrm{shot}} = \frac{2M - x^2}{\varepsilon^2} \label{eq:arbitrary_simultaneously_number_of_shots}
\end{align}
shots.

\subsection{Prefix-product array}\label{sec:level2:PHT_prefix_product_array}

As described in \refsec{sec:level2:qpe_on_early_ftqc}, 
QCELS, a QPE method specialized for Early-FTQC, 
requires the estimation of quantities of the form $\ev{U^m}{\psi}$ for $m = 1, 2, \ldots, M$. 
More generally, we consider the task of estimating $M$ quantities of the form $\ev{U_1}{\psi}, \ev{U_{2:1}}{\psi}, \ldots, \ev{U_{M:1}}{\psi}$, where $U_{b:a}\equiv U_b U_{b-1} \cdots U_a$.
If the $V_{+W}$ gate were used directly for this problem, one would need separate controlled-$U_1$ through controlled-$U_{M:1}$ gates.
Instead, by modifying the ancilla state, the same quantities can be estimated by arranging only one controlled-$U_m$ gate for each $m = 1,\ldots,M$ in sequence, 
which significantly reduces the circuit depth.
Instead of the W state~\eqref{eq:W_state}, an equal superposition of one-hot basis states, 
we use the weight-ladder state $\ket{WL}$ defined by
\begin{align} 
  \ket{WL} &= \frac{1}{\sqrt{M}}\qty(\ket{100\dots0} + \ket{110\dots 0} + \cdots + \ket{111\dots 1})\nonumber\\
          &= \frac{1}{\sqrt{M}}\sum_{m=1}^{M} \ket{e_{1:m}}_M,\label{eq:WL_state}
\end{align}
where $\ket{e_{1:m}}_M$ is the weight-ladder basis state
\begin{align} 
  \ket{e_{1:m}}_M &\equiv \ket{1}^{\otimes m} \otimes \ket{0}^{\otimes (M-m)} \quad \text{for $m = 1, 2, \ldots, M$}.
\end{align}
As in the case of an arbitrary set of unitary operators, we use the statistically optimal choice $\alpha=\beta=1/\sqrt{2}$ and define
\begin{align} 
  \ket{+WL} &\equiv \frac{1}{\sqrt{2}}\qty(\ket{0}^{\otimes M} + \ket{WL}).\label{eq:plus_WL_state}
\end{align}

\begin{figure*}[bt]
\includegraphics[width = 0.65\textwidth]{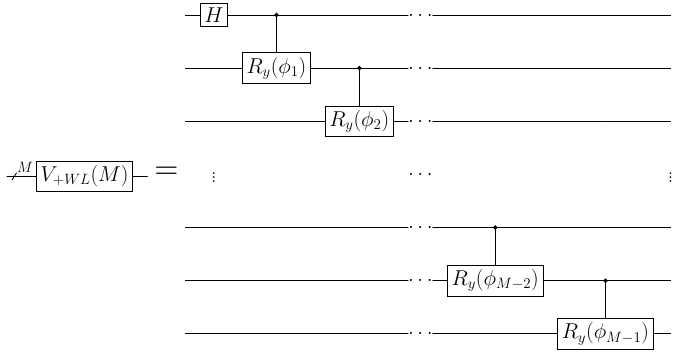}
\caption{\label{fig:V_plusWL} 
Quantum circuit implementing the $V_{+WL}(M)$ gate.
Here, $\phi_i = 2\arccos\qty(\sqrt{\frac{1}{M - i + 1}})$.
}
\end{figure*}

We denote by $V_{+WL}$ the gate that prepares $\ket{+WL}$ from the all-zero state:
\begin{align}
  V_{+WL}(M)\ket{0}^{\otimes M} = \ket{+WL}.
\end{align}
After applying the $V_{+WL}(M)$ gate, the full quantum state becomes
\begin{align}
    &\frac{1}{\sqrt{2}}\qty(\ket{0}^{\otimes M} + \ket{WL}) \otimes \ket{\psi}\nonumber\\
  = &\frac{1}{\sqrt{2}}\ket{0}^{\otimes M} \otimes \ket{\psi} + \frac{1}{\sqrt{2M}} \sum_{m=1}^{M} \ket{e_{1:m}}_M \otimes \ket{\psi}.
\end{align}
After the controlled unitary gates are applied, the full quantum state $\ket{\Psi}$ becomes
\begin{align}
    \ket{\Psi} = \frac{1}{\sqrt{2}}\ket{0}^{\otimes M} \otimes \ket{\psi} + \frac{1}{\sqrt{2M}} \sum_{m=1}^{M} \ket{e_{1:m}}_M \otimes U_{m:1} \ket{\psi}.
\end{align}
When the first $m$ ancilla qubits are measured in the Pauli $X$ basis,
the expectation value of $\otimes_{i=1}^{m} X_i$ is
\begin{align}
  \ev**{\otimes_{i=1}^{m} X_i}{\Psi} &= \frac{1}{\sqrt{M}}\Re \qty[\ev**{U_{m:1}}{\psi}] \nonumber\\
  &\text{for $m = 1, 2, \dots, M$}.
\end{align}
Thus, the real part of $\ev**{U_{m:1}}{\psi}$ is obtained as
\begin{align}
  \Re \qty[\ev**{U_{m:1}}{\psi}] = \sqrt{M} \ev*{\otimes_{i=1}^{m} X_i} \quad \text{for $m = 1, 2, \dots, M$}.
\end{align}
Similarly, if the first ancilla qubit is measured in the Pauli $Y$ basis and the remaining $m-1$ ancilla qubits are measured in the Pauli $X$ basis, then
\begin{align}
  \ev**{Y_1 \otimes \left(\otimes_{i=2}^{m} X_i\right)}{\Psi} &= \frac{1}{\sqrt{M}}\Im \qty[\ev**{U_{m:1}}{\psi}] \nonumber \\
  & \text{for $m = 2, 3, \dots, M$}.
\end{align}
The imaginary part is obtained by multiplying this expectation value by $\sqrt{M}$.
The $V_{+WL}(M)$ gate can be constructed as shown in \reffig{fig:V_plusWL}.
The statistical error and the number of shots are given by the same expressions 
as \refeq{eq:arbitrary_statistical_error} and \refeq{eq:arbitrary_number_of_shots}, respectively, 
for the case of an arbitrary set of unitary operators.

In the present case, the circuit depth can be kept only slightly larger than the maximum depth originally required. 
When using HT to compute $\ev**{U_{M:1}}{\psi}$, it is necessary to arrange all the controlled gates from $U_1$ through $U_M$,
and the same applies to PHT, with the depth increasing by the amount corresponding to the $V$ gate. 
The longer each $U_i$ is, the smaller this additional impact becomes.

\begin{figure*}[bt]
\centering
\includegraphics[width = 0.6\textwidth]{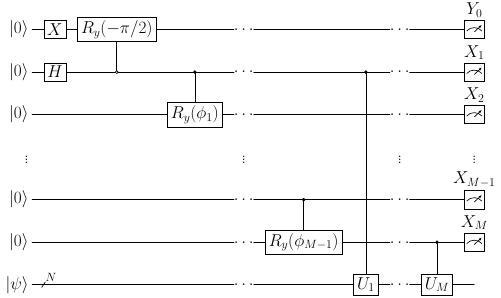}
\caption{
PHT circuit for simultaneously estimating the real and imaginary parts of $\ev**{U_{m:1}}{\psi}$ for $m = 1, 2, \dots, M$.
Here, $\phi_i = 2\arccos\qty(\sqrt{\frac{1}{M - i + 1}})$, and the open circle denotes a negative control.
}
\label{fig:prefix_simultaneously}
\end{figure*}

Moreover, PHT can estimate both the real and imaginary parts of $\ev**{U_{m:1}}{\psi}$ for all $m = 1, 2, \dots, M$ simultaneously with a single circuit type.
The circuit shown in \reffig{fig:prefix_simultaneously} uses one additional ancilla qubit.
The real and imaginary parts are obtained from the corresponding ancilla observables as
\begin{align}
  \Re \qty[\ev**{U_{m:1}}{\psi}] &= \sqrt{2M} \ev{\otimes_{i=1}^{m} X_i}, \label{eq:prefix_simultaneously_real} \\
  \Im \qty[\ev**{U_{m:1}}{\psi}] &= \sqrt{2M} \ev{Y_0 \otimes \left( \otimes_{i=1}^{m} X_i \right)} \label{eq:prefix_simultaneously_imaginary} \\
  & \text{for $m = 1, 2, \dots, M$}. \nonumber
\end{align}
The statistical error and the number of shots are given by the same expressions 
as \refeq{eq:arbitrary_simultaneously_statistical_error} and \refeq{eq:arbitrary_simultaneously_number_of_shots}, respectively, 
for the case of an arbitrary set of unitary operators.

\subsection{Gram-matrix elements}\label{sec:level2:PHT_gram_matrix}

As described in \refsec{sec:level2:quantum_kernel_method}, 
quantum kernel methods require the estimation of quantities of the form $\ev{U_i^\dagger U_j}{\psi}$ for $1\le i<j\le M$. 
As in the prefix-product case, a slight modification of the ancilla state can reduce the circuit depth compared with the construction in \refsec{sec:level2:PHT_arbitrary_set_of_unitary_operators}.
Here, instead of the $\ket{+W}$ state, we use the W state $\ket{W}$ itself.
We denote by $V_W$ the gate that prepares $\ket{W}$ from the all-zero state:
\begin{align}
  V_{W}(M)\ket{0}^{\otimes M} = \ket{W}.
\end{align}
This preparation is obtained from the construction of $V_{+W}(M)$ by setting $\alpha=0$ and $\beta=1$; in the circuit, this corresponds to replacing the initial Hadamard gate by an $X$ gate.
After applying the $V_{W}(M)$ gate, the full quantum state becomes
\begin{align}
    \ket{W} \otimes \ket{\psi}
  = \frac{1}{\sqrt{M}} \sum_{m=1}^{M} \ket{e_{m}}_M \otimes \ket{\psi}.
\end{align}
After the controlled unitary gates are applied, the full quantum state $\ket{\Psi}$ becomes
\begin{align}
    \ket{\Psi} = \frac{1}{\sqrt{M}} \sum_{m=1}^{M} \ket{e_{m}}_M \otimes U_{m} \ket{\psi}.
\end{align}
When the $i$-th and $j$-th ancilla qubits are measured in the Pauli $X$ basis for $1\le i<j\le M$,
the expectation value of $X_i X_j$ is
\begin{align}
    \ev**{X_i X_j}{\Psi}
  &= \frac{1}{M} \qty(\ev**{U_i^\dagger U_j}{\psi} + \ev**{U_j^\dagger U_i}{\psi})\nonumber\\
  &= \frac{2}{M} \Re \qty[\ev**{U_i^\dagger U_j}{\psi}]\nonumber\\
  &\text{for } 1\le i<j\le M.
\end{align}

Thus, the real part of $\ev**{U_i^\dagger U_j}{\psi}$ is obtained as
\begin{align}
  \Re \qty[\ev**{U_i^\dagger U_j}{\psi}] = \frac{M}{2} \ev{X_i X_j} \quad \text{for } 1\le i<j\le M.\label{eq:PHT_general_U_i_dagger_U_j}
\end{align}
The statistical error after $N_{\mathrm{shot}}$ shots is
\begin{align}
  \sigma = \sqrt{\frac{M^2 - 4 x^2}{4N_{\mathrm{shot}}}}. \label{eq:Gram_statistical_error}
\end{align}
Thus, to achieve target precision $\varepsilon$, 
we need
\begin{align}
    N_{\mathrm{shot}} = \frac{M^2 - 4 x^2}{4\varepsilon^2} \label{eq:Gram_number_of_shots}
\end{align}
shots.

Similarly, measuring the $i$-th ancilla qubit in the Pauli $X$ basis and the $j$-th ancilla qubit in the Pauli $Y$ basis gives the imaginary part:
\begin{align}
  \Im \qty[\ev**{U_i^\dagger U_j}{\psi}] = \frac{M}{2} \ev{X_i Y_j} \quad \text{for } 1\le i<j\le M.
\end{align}

When using HT, the combinations of two unitary operators selected from $M$ unitary operators require
$\binom{M}{2}=M(M-1)/2$ distinct quantum circuits for either the real or imaginary part alone.
Since each HT circuit involves two unitary operators, the total number of unitary-operator applications over all such circuits is proportional to $M(M-1)$.
In contrast, the PHT construction applies each of the $M$ unitary operators once, together with the gates required to prepare the $\ket{W}$ state.
Thus, PHT reduces the total circuit workload and the number of distinct circuit types, although the resulting single PHT circuit can be deeper than an individual two-unitary HT circuit.

\begin{figure}[bt]
\centering
\includegraphics[width = 0.45\textwidth]{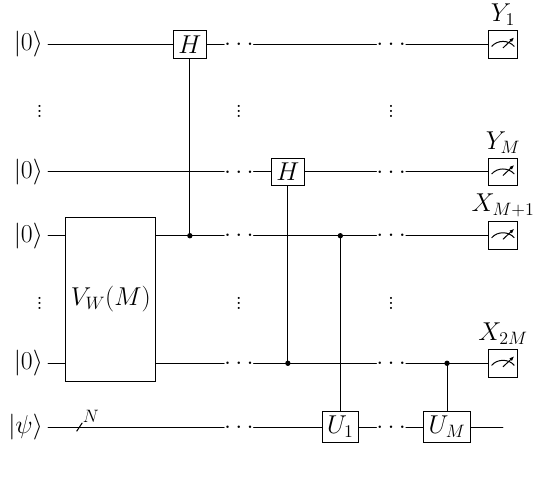}
\caption{PHT circuit for simultaneously estimating the real and imaginary parts of $\ev**{U_i^\dagger U_j}{\psi}$ for $1\le i<j\le M$.
Here, $\phi_i = 2\arccos\qty(\sqrt{\frac{1}{M - i + 1}})$.}
\label{fig:Gram_simultaneously}
\end{figure}

Moreover, PHT can estimate both the real and imaginary parts of $\ev**{U_i^\dagger U_j}{\psi}$ for all $1\le i<j\le M$ simultaneously with a single circuit type. 
The circuit shown in \reffig{fig:Gram_simultaneously} uses $M$ additional ancilla qubits.
The real and imaginary parts are obtained from two equivalent sets of ancilla observables as
\begin{align}
  \Re \qty[\ev**{U_i^\dagger U_j}{\psi}] &= M \ev{X_{M+i} X_{M+j}} \nonumber\\
                                         &= M \ev{Y_i Y_j X_{M+i} X_{M+j}}, \label{eq:Gram_simultaneously_real} \\
  \Im \qty[\ev**{U_i^\dagger U_j}{\psi}] &= - M \ev{Y_i X_{M+i} X_{M+j}}\nonumber\\
                                         &= M \ev{Y_j X_{M+i} X_{M+j}} \label{eq:Gram_simultaneously_imaginary} \\ 
  &\text{for } 1\le i<j\le M. \nonumber
\end{align}
The statistical error after $N_{\mathrm{shot}}$ shots is
\begin{align}
  \sigma = \sqrt{\frac{M^2 - x^2}{N_{\mathrm{shot}}}}.  
\end{align}
Thus, to achieve target precision $\varepsilon$, 
we need
\begin{align}
    N_{\mathrm{shot}} = \frac{M^2 - x^2}{\varepsilon^2}
\end{align}
shots.

To further reduce the statistical error, we can average the two estimators for each of the real and imaginary parts:
\begin{align}
  &\Re \qty[\ev**{U_i^\dagger U_j}{\psi}] \nonumber\\ 
  & ~~ = \frac{M}{2} \qty(\ev{X_{M+i} X_{M+j}} + \ev{Y_i Y_j X_{M+i} X_{M+j}}),\nonumber\\
  &\Im \qty[\ev**{U_i^\dagger U_j}{\psi}] \nonumber\\
  & ~~ = \frac{M}{2} \qty(-\ev{Y_i X_{M+i} X_{M+j}} + \ev{Y_j X_{M+i} X_{M+j}}).
  \label{eq:Gram_simultaneously_average}
\end{align}
The statistical error after $N_{\mathrm{shot}}$ shots is
\begin{align}
  \sigma = \sqrt{\frac{M^2 - 2 x^2}{2 N_{\mathrm{shot}}}}.
\end{align}
This reduces the statistical error beyond what would be achieved by simply doubling the number of shots.
The additional improvement originates from the negative covariance between the two estimators in \refeq{eq:Gram_simultaneously_average}, 
which follows from $\ev{Y_i Y_j} = 0$ in the present setting.
Thus, to achieve target precision $\varepsilon$, 
we need
\begin{align}
    N_{\mathrm{shot}} = \frac{M^2 - 2 x^2}{2\varepsilon^2} \label{eq:Gram_simultaneously_shot}
\end{align}
shots.

Based on the above analysis, we conclude that PHT can reduce the number of distinct circuit types by a factor of order $M^2$.
This reduction comes with a workload-dependent shot-count trade-off.
In particular, when $x$ is sufficiently small, the total number of shots required by PHT can be smaller than that required by HT, as shown in \refsec{subsec:level3:Gram:ShotOverhead}.

\subsection{Ancilla reduction by mid-circuit measurement and reset}\label{sec:level2:mid_circuit}

\begin{figure*}[bt]
\includegraphics[width = 0.7\textwidth]{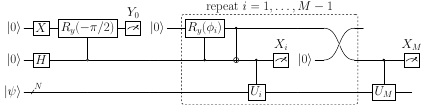}
\caption{\label{fig:arbitrary_simultaneously_mid} 
Quantum circuit obtained from \reffig{fig:arbitrary_simultaneously} by using MCMR, which reduces the number of ancilla qubits to two. 
Here, $\phi_i = 2\arccos\qty(\sqrt{\frac{1}{M - i + 1}})$.
}
\end{figure*}

\begin{figure*}[bt]
\centering
\includegraphics[width = 0.6\textwidth]{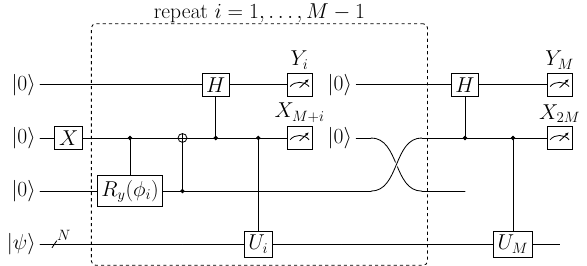}
\caption{\label{fig:Gram_simultaneously_mid} 
Quantum circuit obtained from \reffig{fig:Gram_simultaneously} by using MCMR, which reduces the number of ancilla qubits to three. 
Here, $\phi_i = 2\arccos\qty(\sqrt{\frac{1}{M - i + 1}})$.
}
\end{figure*}

MCMR enables qubits to be measured, reset, and reused within a quantum circuit.
This capability is particularly important for scalable quantum algorithms and quantum error correction schemes
\cite{shor_scheme_1995,knill_quantum_2005,divincenzo_effective_2007,landahl_faulttolerant_2011,fowler_surface_2012,devitt_quantum_2013,riste_realtime_2020}, 
where ancilla qubits are repeatedly measured and reset to extract error syndromes.
It has also been used for error detection~\cite{urbanek_error_2020}, repeat-until-success circuits~\cite{paetznick_repeatuntilsuccess_2014}, 
and dynamic circuits~\cite{corcoles_exploiting_2021}.
MCMR has already been implemented in multiple quantum hardware platforms, 
such as trapped-ion~\cite{pino_demonstration_2021,yu_insitu_2025}, superconducting~\cite{carreravazquez_combining_2024}, 
and neutral atom~\cite{graham_midcircuit_2023} quantum devices. 

The PHT circuits described in 
\refsec{sec:level2:PHT_arbitrary_set_of_unitary_operators}, 
\refsec{sec:level2:PHT_prefix_product_array}, and
\refsec{sec:level2:PHT_gram_matrix}
require many ancilla qubits in their direct implementations.
When MCMR is available, however, this ancilla overhead can be reduced to only two or three qubits.
To recycle an ancilla qubit via MCMR, suppose that a qubit first plays the role of $q_A$ and is later reused as $q_B$ after measurement and reset.
This reuse is valid only if all gates acting on $q_A$ are scheduled before any gate acting on $q_B$.
Under this time ordering, $q_A$ can be measured and reset before it is needed as $q_B$, 
so the two logical ancilla qubits do not need to exist simultaneously as distinct qubits.
For this reason, the MCMR-based construction must use the sequential ancilla-state preparation circuit, rather than the logarithmic-depth tree-like implementation of the $V$ gate.
Consequently, the PHT circuits can be implemented with two ancilla qubits, as illustrated in \reffig{fig:arbitrary_simultaneously_mid}, 
except for the circuit that simultaneously estimates the real and imaginary parts in the Gram-matrix setting, 
which requires three ancilla qubits as shown in \reffig{fig:Gram_simultaneously_mid}.

\section{Cost Analysis}
\label{sec:level1:CostComparison}

In this section, we compare the costs of HT and PHT.
We organize the comparison according to the structural classes of target quantities.
Specifically, we separately analyze
(i) the cases of arbitrary sets of unitary operators and prefix-product arrays
(\refsec{sec:level2:PHT_arbitrary_set_of_unitary_operators}, \refsec{sec:level2:PHT_prefix_product_array})
and (ii) the Gram-matrix case (\refsec{sec:level2:PHT_gram_matrix}).
The source code used for the numerical analysis in this section is available at~\cite{github_parallel_hadamard_tests}.

\subsection{Arbitrary sets of unitary operators and prefix-product arrays}
\label{sec:level2:ArbitraryPrefix}

We first consider the cases of
arbitrary sets of unitary operators and
prefix-product arrays.
Here, we consider the setting in which either the real or imaginary part of each target quantity is estimated,
as in VQE gradient estimation.
In both cases, the number of target quantities to be estimated is $M$,
and the overall cost structure is essentially identical.
Therefore, we analyze these two cases together.

\subsubsection{Comparison of total cost}
\label{subsec:level3:ArbitraryPrefix:TotalCost}

Let $x$ denote a general expectation value to be estimated.
For simplicity of comparison, we assume throughout this section
that all target quantities have the same value $x$.

In the standard HT-based approach,
one distinct quantum circuit is required for each target quantity.
Thus, the total number of distinct circuit types is $M$.
From the statistical analysis in \refeq{eq:HT_shot},
the number of shots required per circuit to achieve the target precision parameter $\varepsilon$ is
\begin{align}
N_{\mathrm{shot}}^{\mathrm{HT}} = \frac{1 - x^{2}}{\varepsilon^{2}}.
\end{align}
The resulting total cost is therefore
\begin{align}
C_{\mathrm{HT}}
= M C_{\mathrm{task}} + M \frac{1 - x^{2}}{\varepsilon^{2}} C_{\mathrm{shot}}.
\end{align}
Here, $C_{\mathrm{task}}$ denotes the cost per circuit type, and $C_{\mathrm{shot}}$ denotes the cost per shot.
For simplicity, we assume that the per-shot cost is the same for HT and PHT.

In contrast, PHT evaluates all $M$ quantities using a single quantum circuit.
As shown in \refeq{eq:arbitrary_number_of_shots},
the required number of shots to achieve the target precision parameter $\varepsilon$ is
\begin{align}
N_{\mathrm{shot}}^{\mathrm{PHT}} = \frac{M - x^{2}}{\varepsilon^{2}}.
\end{align}
The total cost of PHT is therefore given by
\begin{align}
C_{\mathrm{PHT}}
= C_{\mathrm{task}} + \frac{M - x^{2}}{\varepsilon^{2}} C_{\mathrm{shot}}.
\end{align}

Comparing these two expressions gives
\begin{align}
C_{\mathrm{PHT}} < C_{\mathrm{HT}}
\quad \Longleftrightarrow \quad
\frac{C_{\mathrm{task}}}{C_{\mathrm{shot}}}
> \frac{x^{2}}{\varepsilon^{2}}.
\label{eq:level4:ArbitraryPrefix:Inequality}
\end{align}
Therefore, PHT is advantageous over HT when
\begin{itemize}
  \item the cost per circuit type is sufficiently large relative to the cost per shot,
  \item the target quantities to be estimated are small, and
  \item the required precision is not excessively high.
\end{itemize}

\subsubsection{Numerical analysis}
\label{subsec:level3:ArbitraryPrefix:NumericalAnalysis}

Although a quantitative comparison in the FTQC regime requires many assumptions and is therefore challenging, 
such an analysis is more concrete in the NISQ setting.
We therefore present a comparative study in the context of VQE.
We compare the total costs by substituting concrete numerical values for both time and financial costs.

In all methods, the Hamiltonian must be decomposed into tensor products of Pauli strings,
and the expectation values and gradients must be evaluated for each string individually. 
Therefore, for the purpose of comparison, we focus on a single Pauli string. 
This simplification does not affect the generality of the cost comparison. 
Although commuting Pauli strings can be grouped and measured simultaneously~\cite{verteletskyi_measurement_2020}, 
this technique is applicable to both HT and PHT and is therefore not considered here. 
In addition, a previous study proposed swapping the roles of the gate generator and the observable~\cite{li_efficient_2024}. 
Such techniques do not provide an advantage in our setting, where the generator of the unitary operator is itself unitary, 
and are therefore also excluded from the comparison.

\begin{figure}[bt]
\centering
\includegraphics[width = 0.45\textwidth]{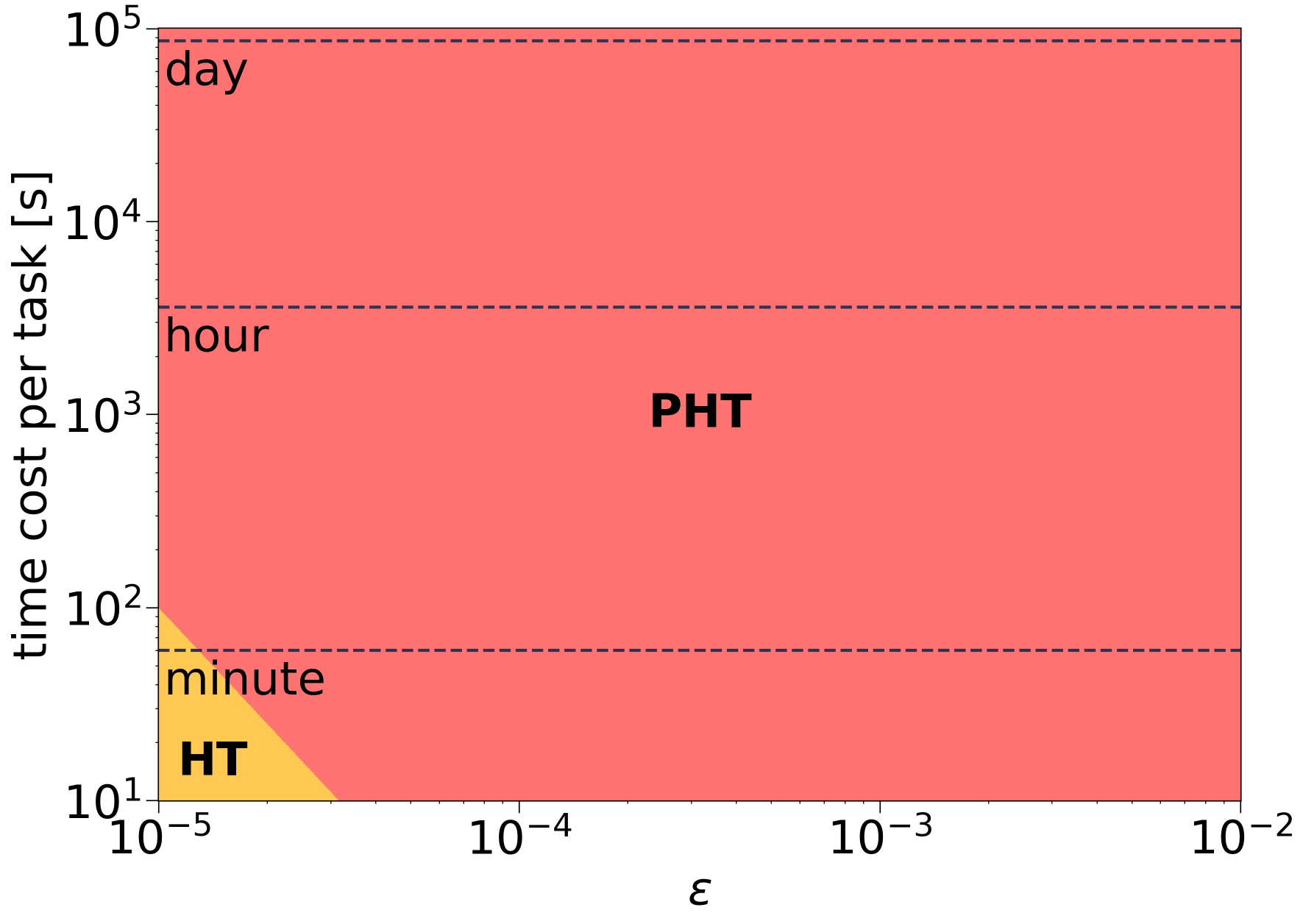}
\caption{\label{fig:result_time}
Cost-optimal method in terms of total time cost.
The dashed horizontal lines indicate one minute, one hour, and one day.
Here, we assume $C_{\mathrm{shot}} = 100 ~\si{\micro \second}$, $M = 5$, 
and $x = 0.01$, where $x$ denotes the true gradient value.
}
\end{figure}

For the time-cost comparison, we fix the time per shot at $\SI{100}{\micro \second}$,
which has been reported as a typical quantum-circuit execution time on IBM Quantum~\cite{ibm_quantum_workload}.
We compute the total cost of each method by varying $\varepsilon$ and $C_{\mathrm{task}}$, 
and use a color map to indicate the lower-cost method in each region, as shown in \reffig{fig:result_time}.
As expected, PHT is favorable when the required precision is not very high and the time cost per circuit type is large. 
In practice, demand for quantum computers has been rapidly increasing, 
whereas the number of cloud services providing access to quantum hardware remains limited. 
Consequently, queueing times often extend from several hours to multiple days, suggesting that PHT can help mitigate this bottleneck.

\begin{figure}[bt]
\centering
\includegraphics[width = 0.45\textwidth]{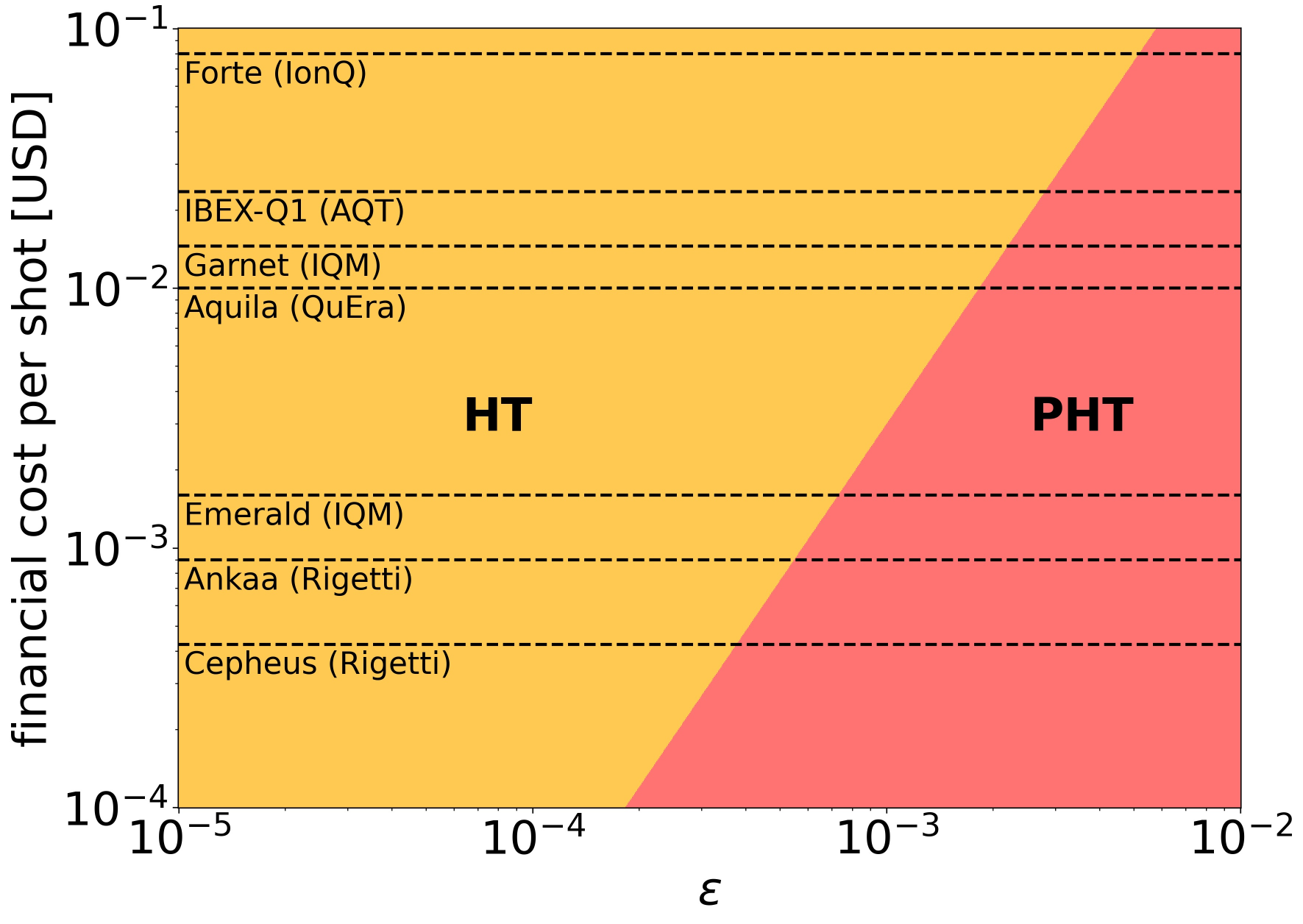}
\caption{\label{fig:result_financial}
Cost-optimal method in terms of total financial cost.
The dashed lines indicate the actual Amazon Braket prices~\protect\cite{aws_braket_pricing} for the devices listed in \reftab{tab:amazon_braket}.
Here, we assume $C_{\mathrm{task}} = 0.3 ~\mathrm{USD}$, $M = 5$, 
and $x = 0.01$, where $x$ denotes the true gradient value.
}
\end{figure}

For the financial-cost comparison, we fix the cost per circuit type at $0.3 ~\mathrm{USD}$,
which is the task price used by Amazon Braket~\cite{aws_braket_pricing} and qBraid~\cite{qbraid_pricing}.
We compute the total cost of each method by varying $\varepsilon$ and $C_{\mathrm{shot}}$, 
and again indicate the lower-cost method in each region using a color map, as shown in \reffig{fig:result_financial}.
Because the per-shot cost depends on the device, 
the actual Amazon Braket prices for each device are shown as dashed lines in the figure.
These prices are summarized in \reftab{tab:amazon_braket}.
As expected, PHT is favorable when the required precision is not very high and the per-shot cost is low. 
In VQE, it is often sufficient to identify parameters with large gradients, 
and highly accurate gradient estimates are not always required at every parameter update. 
This is precisely the type of regime in which PHT is advantageous.
Choosing a device with a low per-shot cost is also important.
However, such devices may have fewer qubits or higher noise levels, so these trade-offs must be taken into account.

\begin{table*}[bt]
   \caption{Financial costs per task (circuit type) and per shot on Amazon Braket~\protect\cite{aws_braket_pricing}, as of June 17, 2026.}
   \label{tab:amazon_braket}
   \centering
   \begin{tabular}{cc|cc}
   \hline \hline
   Hardware Provider & QPU family & Per-task price [USD] & Per-shot price [USD] \\
   \hline
   AQT     & IBEX-Q1 & 0.30000 & 0.02350\\
   IonQ    & Forte   & 0.30000 & 0.08000\\
   IQM     & Emerald & 0.30000 & 0.00160\\
   IQM     & Garnet  & 0.30000 & 0.00145\\
   QuEra   & Aquila  & 0.30000 & 0.01000\\
   Rigetti & Ankaa   & 0.30000 & 0.00090\\
   Rigetti & Cepheus & 0.30000 & 0.000425\\
   \hline \hline
   \end{tabular}
\end{table*}

\subsubsection{Shot overhead}
\label{subsec:level3:ArbitraryPrefix:ShotOverhead}

The difference in the total number of shots between PHT and HT is given by
\begin{align}
\Delta N_{\mathrm{shot}}
:= N_{\mathrm{shot}}^{\mathrm{PHT}} - M N_{\mathrm{shot}}^{\mathrm{HT}}
= \frac{(M-1)x^{2}}{\varepsilon^{2}}.
\end{align}
Thus, PHT requires more shots than HT while drastically reducing the number of distinct circuit types.

\subsubsection{Remarks on variational quantum eigensolver and parameter-shift rule}
\label{subsec:level3:ArbitraryPrefix:VQE}

As mentioned in \refsec{sec:level2:gradient_estimation},
in the context of VQE,
gradient estimation can also be performed using PSR~\cite{mitarai_quantum_2018,schuld_evaluating_2019,hubregtsen_singlecomponent_2022,crooks_gradients_2019,izmaylov_analytic_2021,wierichs_general_2022}.
However, PSR requires two distinct quantum circuits per parameter.
Thus, in terms of the number of circuit types, PSR is less favorable than HT in the present setting.
Since we focus on a regime in which costs that scale with the number of circuit types are significant, 
PSR is also more costly than HT in this comparison.

\subsection{Gram-matrix elements}
\label{sec:level2:Gram}

We next consider the Gram-matrix case introduced in \refsec{sec:level2:PHT_gram_matrix}.
Here, we compare the setting in which both the real and imaginary parts of all target quantities are estimated, 
as required in applications such as quantum kernel methods.

\subsubsection{Comparison of total cost}
\label{subsec:level3:Gram:TotalCost}

In the Gram-matrix setting,
HT requires evaluating all pairs of unitary operators and separately estimating the real and imaginary parts for each pair.
As a result, the number of distinct quantum circuits scales as $M(M-1)$.
The total cost of HT is therefore given by
\begin{align}
C_{\mathrm{HT}}^{\mathrm{Gram}}
= M(M-1) C_{\mathrm{task}}
+ M(M-1) \frac{1 - x^{2}}{\varepsilon^{2}} C_{\mathrm{shot}}.
\end{align}

In PHT, all $M$ unitary operators are applied only once.
As shown in \refeq{eq:Gram_simultaneously_shot}, 
the required number of shots to achieve the target precision parameter $\varepsilon$ is
\begin{align}
N_{\mathrm{shot}}^{\mathrm{PHT,Gram}} = \frac{M^{2} - 2 x^{2}}{2 \varepsilon^{2}}.
\end{align}
The total cost of PHT in this case is therefore given by
\begin{align}
C_{\mathrm{PHT}}^{\mathrm{Gram}}
= C_{\mathrm{task}}
+ \frac{M^{2} - 2x^{2}}{2\varepsilon^{2}} C_{\mathrm{shot}}.
\end{align}

\subsubsection{Shot overhead and reduction}
\label{subsec:level3:Gram:ShotOverhead}

The difference in the total number of shots between PHT and HT is given by
\begin{align}
\Delta N_{\mathrm{shot}}^{\mathrm{Gram}}
&= N_{\mathrm{shot}}^{\mathrm{PHT,Gram}} - M(M-1) N_{\mathrm{shot}}^{\mathrm{HT}}\nonumber\\
&= \frac{2 (M^2 - M - 1) x^2 - M (M - 2)}{2\varepsilon^{2}}.
\end{align}
This expression shows that,
when $x$ is sufficiently small,
the total number of shots required by PHT can be smaller than that required by HT.

To empirically validate the predicted shot reduction in the Gram-matrix setting, 
we construct a toy instance in which all target overlaps are zero. 

\paragraph{Problem setup.}
We consider a two-qubit system and prepare the input state as
\begin{align}
\ket{\psi}=\ket{00}.
\end{align}
We set $M=4$ unitary operators as Pauli-$X$ strings indexed by $\bm{s} = (s_1, s_2) \in\{0,1\}^{2}$:
\begin{align}
U_{\bm{s}} := X^{s_1}\otimes X^{s_2}
\qquad
(\bm{s}\in\{00,01,10,11\}).
\end{align}
Then $U_{\bm{s}}\ket{00}=\ket{\bm{s}}$, and the set $\{\ket{\bm{s}}\}$ forms an orthonormal computational basis. Therefore, for any distinct pair $\bm{s}_i\neq \bm{s}_j$,
\begin{align}
\ev{U_{\bm{s}_i}^{\dagger}U_{\bm{s}_j}}{\psi}
=
\braket{\bm{s}_i}{\bm{s}_j}
=0,
\end{align}
and hence all off-diagonal Gram-matrix elements are exactly zero (and purely real). 
In this experiment, we estimate the real and imaginary parts of $\ev{U_i^{\dagger}U_j}{\psi}$ 
for all pairs $(i,j)$ with $i<j$, resulting in $2 \binom{M}{2} = 12$ target quantities, all with the true value $x=0$.

\paragraph{Shot allocation and expected scaling.}
For HT, each pair $(i,j)$ is estimated by a distinct quantum circuit with $N_{\mathrm{shot}}^{\mathrm{HT}} = 1/\varepsilon^{2}$ shots, so the total shot count scales as
\begin{align}
N_{\mathrm{tot}}^{\mathrm{HT}} = 2\binom{M}{2}\frac{1}{\varepsilon^{2}} = \frac{12}{\varepsilon^{2}}.
\end{align}
For PHT in the Gram-matrix setting, where both the real and imaginary parts are estimated simultaneously,
all quantities are estimated using a single circuit type, and the required number of shots is
\begin{align}
N_{\mathrm{tot}}^{\mathrm{PHT}} = \frac{M^{2}}{2\varepsilon^{2}}=\frac{8}{\varepsilon^{2}},
\end{align}
which is equal to $2/3$ of $N_{\mathrm{tot}}^{\mathrm{HT}}$.

\begin{figure}[bt]
\centering
\includegraphics[width=0.95\linewidth]{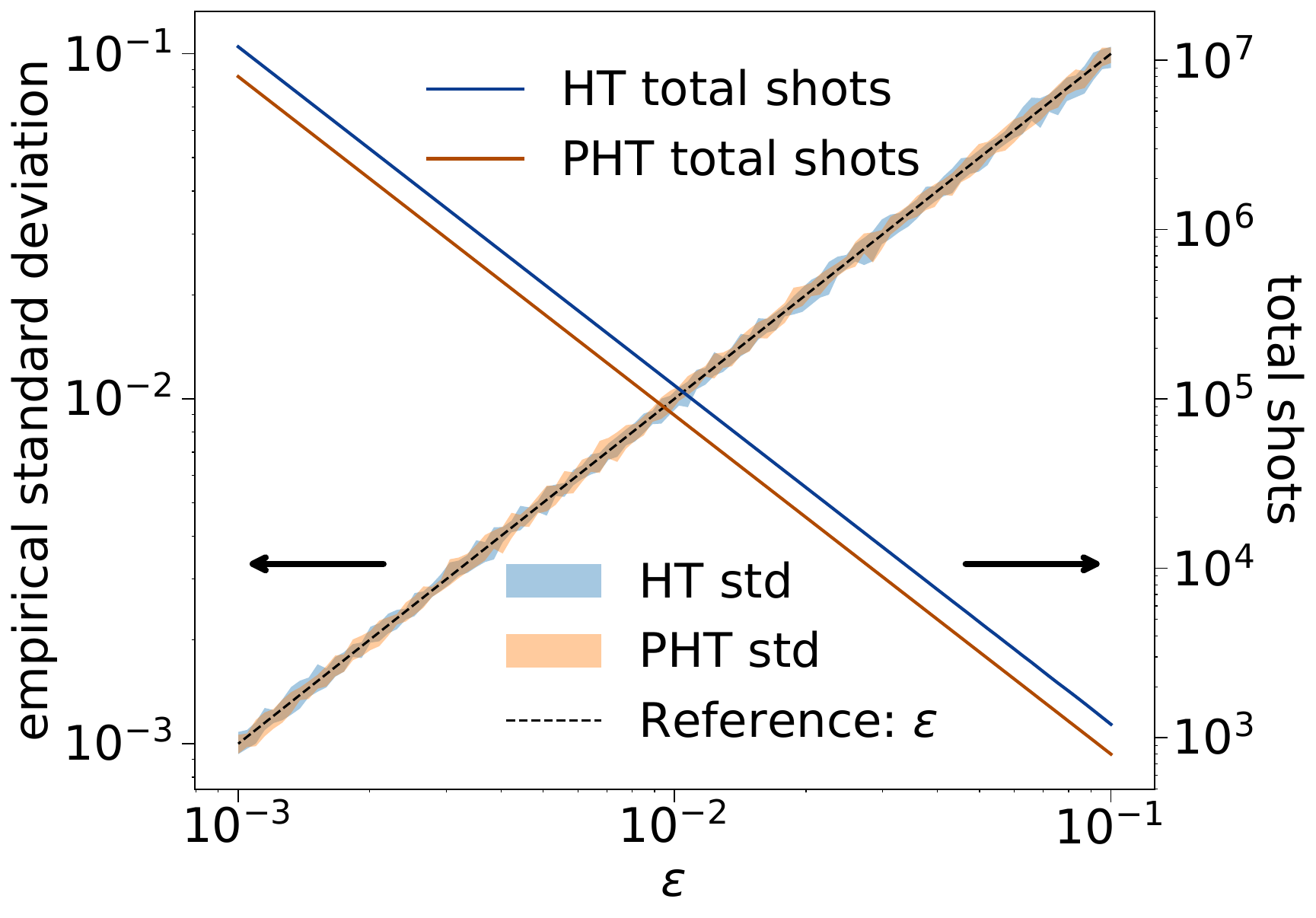}
\caption{
Empirical standard deviation and total number of shots for estimating the off-diagonal Gram-matrix elements
$\ev{U_i^\dagger U_j}{\psi}$ ($i<j$) using HT and PHT.
The horizontal axis shows the target precision parameter $\varepsilon$.
The shaded bands should be read against the left vertical axis and show the min--max range of the empirical standard deviation over twelve quantities, corresponding to the real and imaginary parts of the six off-diagonal pairs.
For each value of $\varepsilon$, the empirical standard deviation is computed from 300 independent simulations with different random seeds.
The dashed black line indicates the target standard deviation $\varepsilon$; 
the number of shots is chosen so that the theoretical standard deviation equals this value.
The solid lines should be read against the right vertical axis and show the total number of shots required by each method.
The results show that PHT can reduce the required number of shots while maintaining the same estimation accuracy as HT.
}
\label{fig:Gram:shot_reduction}
\end{figure}

\paragraph{Numerical verification (Qiskit Aer).}
We simulate both HT and PHT circuits using Qiskit Aer.
For each value of the target precision parameter $\varepsilon$, 
we set the number of shots according to the formulas above and repeat the estimation procedure with independent random seeds to empirically evaluate the standard deviation of each estimate. 
In \reffig{fig:Gram:shot_reduction}, we vary $\varepsilon$ and report, for each method, the total number of shots used and the resulting standard deviation. 
The figure shows that PHT achieves the same estimation error while using fewer shots.
These results provide a concrete demonstration that, in the small-$x$ regime, PHT can reduce not only the number of circuit types but also the total number of shots in the Gram-matrix setting.

\section{Discussion and Conclusion}\label{sec:level1:discussion_and_conclusion}

In this work, we introduced PHT, a method for parallelizing multiple HTs within a single quantum circuit.
PHT is designed to reduce the operational and financial overheads associated with many-instance HTs 
in practical quantum workloads such as VQE, QCELS, and quantum kernel methods.
The method uses a specially prepared ancilla superposition state 
and sequential controlled applications of the target unitary operators.
When MCMR is available, the required number of ancilla qubits can be reduced to two or three.
We provided explicit constructions of the ancilla preparation circuits 
and derived estimator relations for extracting the desired quantities from measured ancilla observables.

Analytically, we characterized the central trade-off of PHT:
it replaces the multiplicity of distinct circuit types with an increase in the single-shot variance of each individual estimate.
For arbitrary sets of unitary operators and prefix-product arrays, this leads to a larger total number of shots.
In the Gram-matrix case, however, the total number of shots can be reduced when the typical off-diagonal overlaps are sufficiently small.
We compared HT and PHT according to the structural classes,
arbitrary sets of unitary operators, prefix-product arrays, and Gram-matrix elements,
under the simplifying assumption that the target quantities share a common typical magnitude.
From this analysis, we identified regimes where PHT is advantageous.
These regimes occur principally when per-circuit fixed costs (compilation, queueing, or cloud per-task charges) dominate the total cost,
when the required precision is modest, or when the true values to be estimated are small.
In the Gram-matrix setting, the number of circuit types required by HT grows rapidly with the problem size,
which makes PHT particularly attractive and can even reduce the overall number of shots when typical overlaps are small.

There are practical limitations relevant to near-term and early fault-tolerant devices. 
Concerns include the accumulation of errors when quantum circuits become deeper
as well as constraints imposed by device connectivity and the availability of reliable MCMR.
These hardware-dependent effects couple nontrivially with the statistical trade-off, 
so the realized benefit of PHT will depend strongly on device noise characteristics and on the effectiveness of compilation
for the resulting circuits.

Overall, PHT provides a promising approach to mitigating the overhead incurred by executing large numbers of HTs in practical quantum workloads.
Although the increase in single-shot variance must be taken into account,
the total shot count depends on the workload structure and can even decrease in the Gram-matrix setting when typical overlaps are small.
PHT therefore offers a new point in the resource trade-off space that can be preferable in cloud-based and throughput-constrained settings.
Future work should focus on benchmarking on real quantum hardware, 
incorporating realistic noise models into cost analyses,
and developing methods that automatically select measurement strategies according to device performance and user cost constraints.
Another important direction is to explore circuit reduction techniques for algorithms beyond HT, 
thereby broadening the applicability of resource-efficient quantum computation.

\begin{acknowledgments}
We thank Kazuki Inomata and Keisuke Murota for fruitful discussions and comments.
This work was supported by the Center of Innovation for Sustainable Quantum AI (JST Grant Number JPMJPF2221).
S.I. acknowledges support from FoPM, a WINGS Program of the University of Tokyo.
\end{acknowledgments}

\bibliography{paper_PHT}

\begin{thebibliography}{72}%
\makeatletter
\providecommand \@ifxundefined [1]{%
 \@ifx{#1\undefined}
}%
\providecommand \@ifnum [1]{%
 \ifnum #1\expandafter \@firstoftwo
 \else \expandafter \@secondoftwo
 \fi
}%
\providecommand \@ifx [1]{%
 \ifx #1\expandafter \@firstoftwo
 \else \expandafter \@secondoftwo
 \fi
}%
\providecommand \natexlab [1]{#1}%
\providecommand \enquote  [1]{``#1''}%
\providecommand \bibnamefont  [1]{#1}%
\providecommand \bibfnamefont [1]{#1}%
\providecommand \citenamefont [1]{#1}%
\providecommand \href@noop [0]{\@secondoftwo}%
\providecommand \href [0]{\begingroup \@sanitize@url \@href}%
\providecommand \@href[1]{\@@startlink{#1}\@@href}%
\providecommand \@@href[1]{\endgroup#1\@@endlink}%
\providecommand \@sanitize@url [0]{\catcode `\\12\catcode `\$12\catcode `\&12\catcode `\#12\catcode `\^12\catcode `\_12\catcode `\%12\relax}%
\providecommand \@@startlink[1]{}%
\providecommand \@@endlink[0]{}%
\providecommand \url  [0]{\begingroup\@sanitize@url \@url }%
\providecommand \@url [1]{\endgroup\@href {#1}{\urlprefix }}%
\providecommand \urlprefix  [0]{URL }%
\providecommand \Eprint [0]{\href }%
\providecommand \doibase [0]{https://doi.org/}%
\providecommand \selectlanguage [0]{\@gobble}%
\providecommand \bibinfo  [0]{\@secondoftwo}%
\providecommand \bibfield  [0]{\@secondoftwo}%
\providecommand \translation [1]{[#1]}%
\providecommand \BibitemOpen [0]{}%
\providecommand \bibitemStop [0]{}%
\providecommand \bibitemNoStop [0]{.\EOS\space}%
\providecommand \EOS [0]{\spacefactor3000\relax}%
\providecommand \BibitemShut  [1]{\csname bibitem#1\endcsname}%
\let\auto@bib@innerbib\@empty
\bibitem [{\citenamefont {Shor}(1994)}]{shor_algorithms_1994}%
  \BibitemOpen
  \bibfield  {author} {\bibinfo {author} {\bibfnamefont {P.~W.}\ \bibnamefont {Shor}},\ }\bibfield  {title} {\bibinfo {title} {Algorithms for quantum computation: Discrete logarithms and factoring},\ }in\ \href {https://doi.org/10.1109/SFCS.1994.365700} {\emph {\bibinfo {booktitle} {Proceedings 35th {{Annual Symposium}} on {{Foundations}} of {{Computer Science}}}}}\ (\bibinfo {year} {1994})\ pp.\ \bibinfo {pages} {124--134}\BibitemShut {NoStop}%
\bibitem [{\citenamefont {Shor}(1999)}]{shor_polynomialtime_1999}%
  \BibitemOpen
  \bibfield  {author} {\bibinfo {author} {\bibfnamefont {P.~W.}\ \bibnamefont {Shor}},\ }\bibfield  {title} {\bibinfo {title} {Polynomial-{{Time Algorithms}} for {{Prime Factorization}} and {{Discrete Logarithms}} on a {{Quantum Computer}}},\ }\href {https://doi.org/10.1137/S0036144598347011} {\bibfield  {journal} {\bibinfo  {journal} {SIAM Rev.}\ }\textbf {\bibinfo {volume} {41}},\ \bibinfo {pages} {303} (\bibinfo {year} {1999})}\BibitemShut {NoStop}%
\bibitem [{\citenamefont {Grover}(1996)}]{grover_fast_1996}%
  \BibitemOpen
  \bibfield  {author} {\bibinfo {author} {\bibfnamefont {L.~K.}\ \bibnamefont {Grover}},\ }\bibfield  {title} {\bibinfo {title} {A fast quantum mechanical algorithm for database search},\ }in\ \href {https://doi.org/10.1145/237814.237866} {\emph {\bibinfo {booktitle} {Proceedings of the Twenty-Eighth Annual {{ACM}} Symposium on {{Theory}} of Computing - {{STOC}} '96}}}\ (\bibinfo {address} {Philadelphia, Pennsylvania, United States},\ \bibinfo {year} {1996})\ pp.\ \bibinfo {pages} {212--219}\BibitemShut {NoStop}%
\bibitem [{\citenamefont {Grover}(1997)}]{grover_quantum_1997}%
  \BibitemOpen
  \bibfield  {author} {\bibinfo {author} {\bibfnamefont {L.~K.}\ \bibnamefont {Grover}},\ }\bibfield  {title} {\bibinfo {title} {Quantum {{Mechanics Helps}} in {{Searching}} for a {{Needle}} in a {{Haystack}}},\ }\href {https://doi.org/10.1103/PhysRevLett.79.325} {\bibfield  {journal} {\bibinfo  {journal} {Phys. Rev. Lett.}\ }\textbf {\bibinfo {volume} {79}},\ \bibinfo {pages} {325} (\bibinfo {year} {1997})}\BibitemShut {NoStop}%
\bibitem [{\citenamefont {Lloyd}(1996)}]{lloyd_universal_1996}%
  \BibitemOpen
  \bibfield  {author} {\bibinfo {author} {\bibfnamefont {S.}~\bibnamefont {Lloyd}},\ }\bibfield  {title} {\bibinfo {title} {Universal {{Quantum Simulators}}},\ }\href {https://doi.org/10.1126/science.273.5278.1073} {\bibfield  {journal} {\bibinfo  {journal} {Science}\ }\textbf {\bibinfo {volume} {273}},\ \bibinfo {pages} {1073} (\bibinfo {year} {1996})}\BibitemShut {NoStop}%
\bibitem [{\citenamefont {Abrams}\ and\ \citenamefont {Lloyd}(1999)}]{abrams_quantum_1999}%
  \BibitemOpen
  \bibfield  {author} {\bibinfo {author} {\bibfnamefont {D.~S.}\ \bibnamefont {Abrams}}\ and\ \bibinfo {author} {\bibfnamefont {S.}~\bibnamefont {Lloyd}},\ }\bibfield  {title} {\bibinfo {title} {Quantum {{Algorithm Providing Exponential Speed Increase}} for {{Finding Eigenvalues}} and {{Eigenvectors}}},\ }\href {https://doi.org/10.1103/PhysRevLett.83.5162} {\bibfield  {journal} {\bibinfo  {journal} {Phys. Rev. Lett.}\ }\textbf {\bibinfo {volume} {83}},\ \bibinfo {pages} {5162} (\bibinfo {year} {1999})}\BibitemShut {NoStop}%
\bibitem [{\citenamefont {{Aspuru-Guzik}}\ \emph {et~al.}(2005)\citenamefont {{Aspuru-Guzik}}, \citenamefont {Dutoi}, \citenamefont {Love},\ and\ \citenamefont {{Head-Gordon}}}]{aspuru-guzik_simulated_2005}%
  \BibitemOpen
  \bibfield  {author} {\bibinfo {author} {\bibfnamefont {A.}~\bibnamefont {{Aspuru-Guzik}}}, \bibinfo {author} {\bibfnamefont {A.~D.}\ \bibnamefont {Dutoi}}, \bibinfo {author} {\bibfnamefont {P.~J.}\ \bibnamefont {Love}},\ and\ \bibinfo {author} {\bibfnamefont {M.}~\bibnamefont {{Head-Gordon}}},\ }\bibfield  {title} {\bibinfo {title} {Simulated {{Quantum Computation}} of {{Molecular Energies}}},\ }\href {https://doi.org/10.1126/science.1113479} {\bibfield  {journal} {\bibinfo  {journal} {Science}\ }\textbf {\bibinfo {volume} {309}},\ \bibinfo {pages} {1704} (\bibinfo {year} {2005})}\BibitemShut {NoStop}%
\bibitem [{\citenamefont {Harrow}\ \emph {et~al.}(2009)\citenamefont {Harrow}, \citenamefont {Hassidim},\ and\ \citenamefont {Lloyd}}]{harrow_quantum_2009}%
  \BibitemOpen
  \bibfield  {author} {\bibinfo {author} {\bibfnamefont {A.~W.}\ \bibnamefont {Harrow}}, \bibinfo {author} {\bibfnamefont {A.}~\bibnamefont {Hassidim}},\ and\ \bibinfo {author} {\bibfnamefont {S.}~\bibnamefont {Lloyd}},\ }\bibfield  {title} {\bibinfo {title} {Quantum {{Algorithm}} for {{Linear Systems}} of {{Equations}}},\ }\href {https://doi.org/10.1103/PhysRevLett.103.150502} {\bibfield  {journal} {\bibinfo  {journal} {Phys. Rev. Lett.}\ }\textbf {\bibinfo {volume} {103}},\ \bibinfo {pages} {150502} (\bibinfo {year} {2009})}\BibitemShut {NoStop}%
\bibitem [{\citenamefont {Wiebe}\ \emph {et~al.}(2012)\citenamefont {Wiebe}, \citenamefont {Braun},\ and\ \citenamefont {Lloyd}}]{wiebe_quantum_2012}%
  \BibitemOpen
  \bibfield  {author} {\bibinfo {author} {\bibfnamefont {N.}~\bibnamefont {Wiebe}}, \bibinfo {author} {\bibfnamefont {D.}~\bibnamefont {Braun}},\ and\ \bibinfo {author} {\bibfnamefont {S.}~\bibnamefont {Lloyd}},\ }\bibfield  {title} {\bibinfo {title} {Quantum {{Algorithm}} for {{Data Fitting}}},\ }\href {https://doi.org/10.1103/PhysRevLett.109.050505} {\bibfield  {journal} {\bibinfo  {journal} {Phys. Rev. Lett.}\ }\textbf {\bibinfo {volume} {109}},\ \bibinfo {pages} {050505} (\bibinfo {year} {2012})}\BibitemShut {NoStop}%
\bibitem [{\citenamefont {Rebentrost}\ \emph {et~al.}(2014)\citenamefont {Rebentrost}, \citenamefont {Mohseni},\ and\ \citenamefont {Lloyd}}]{rebentrost_quantum_2014}%
  \BibitemOpen
  \bibfield  {author} {\bibinfo {author} {\bibfnamefont {P.}~\bibnamefont {Rebentrost}}, \bibinfo {author} {\bibfnamefont {M.}~\bibnamefont {Mohseni}},\ and\ \bibinfo {author} {\bibfnamefont {S.}~\bibnamefont {Lloyd}},\ }\bibfield  {title} {\bibinfo {title} {Quantum {{Support Vector Machine}} for {{Big Data Classification}}},\ }\href {https://doi.org/10.1103/PhysRevLett.113.130503} {\bibfield  {journal} {\bibinfo  {journal} {Phys. Rev. Lett.}\ }\textbf {\bibinfo {volume} {113}},\ \bibinfo {pages} {130503} (\bibinfo {year} {2014})}\BibitemShut {NoStop}%
\bibitem [{\citenamefont {Cao}\ \emph {et~al.}(2017)\citenamefont {Cao}, \citenamefont {Guerreschi},\ and\ \citenamefont {{Aspuru-Guzik}}}]{cao_quantum_2017}%
  \BibitemOpen
  \bibfield  {author} {\bibinfo {author} {\bibfnamefont {Y.}~\bibnamefont {Cao}}, \bibinfo {author} {\bibfnamefont {G.~G.}\ \bibnamefont {Guerreschi}},\ and\ \bibinfo {author} {\bibfnamefont {A.}~\bibnamefont {{Aspuru-Guzik}}},\ }\bibfield  {title} {\bibinfo {title} {Quantum {{Neuron}}: An elementary building block for machine learning on quantum computers},\ }\bibfield  {journal} {\bibinfo  {journal} {arXiv preprint}\ }\href {https://doi.org/10.48550/arXiv.1711.11240} {10.48550/arXiv.1711.11240} (\bibinfo {year} {2017}),\ \Eprint {https://arxiv.org/abs/1711.11240} {arXiv:1711.11240} \BibitemShut {NoStop}%
\bibitem [{\citenamefont {Kerenidis}\ and\ \citenamefont {Prakash}(2020)}]{kerenidis_quantum_2020}%
  \BibitemOpen
  \bibfield  {author} {\bibinfo {author} {\bibfnamefont {I.}~\bibnamefont {Kerenidis}}\ and\ \bibinfo {author} {\bibfnamefont {A.}~\bibnamefont {Prakash}},\ }\bibfield  {title} {\bibinfo {title} {Quantum gradient descent for linear systems and least squares},\ }\href {https://doi.org/10.1103/PhysRevA.101.022316} {\bibfield  {journal} {\bibinfo  {journal} {Phys. Rev. A}\ }\textbf {\bibinfo {volume} {101}},\ \bibinfo {pages} {022316} (\bibinfo {year} {2020})}\BibitemShut {NoStop}%
\bibitem [{\citenamefont {Cerezo}\ \emph {et~al.}(2021)\citenamefont {Cerezo}, \citenamefont {Arrasmith}, \citenamefont {Babbush}, \citenamefont {Benjamin}, \citenamefont {Endo}, \citenamefont {Fujii}, \citenamefont {McClean}, \citenamefont {Mitarai}, \citenamefont {Yuan}, \citenamefont {Cincio},\ and\ \citenamefont {Coles}}]{cerezo_variational_2021}%
  \BibitemOpen
  \bibfield  {author} {\bibinfo {author} {\bibfnamefont {M.}~\bibnamefont {Cerezo}}, \bibinfo {author} {\bibfnamefont {A.}~\bibnamefont {Arrasmith}}, \bibinfo {author} {\bibfnamefont {R.}~\bibnamefont {Babbush}}, \bibinfo {author} {\bibfnamefont {S.~C.}\ \bibnamefont {Benjamin}}, \bibinfo {author} {\bibfnamefont {S.}~\bibnamefont {Endo}}, \bibinfo {author} {\bibfnamefont {K.}~\bibnamefont {Fujii}}, \bibinfo {author} {\bibfnamefont {J.~R.}\ \bibnamefont {McClean}}, \bibinfo {author} {\bibfnamefont {K.}~\bibnamefont {Mitarai}}, \bibinfo {author} {\bibfnamefont {X.}~\bibnamefont {Yuan}}, \bibinfo {author} {\bibfnamefont {L.}~\bibnamefont {Cincio}},\ and\ \bibinfo {author} {\bibfnamefont {P.~J.}\ \bibnamefont {Coles}},\ }\bibfield  {title} {\bibinfo {title} {Variational quantum algorithms},\ }\href {https://doi.org/10.1038/s42254-021-00348-9} {\bibfield  {journal} {\bibinfo  {journal} {Nat. Rev. Phys.}\ }\textbf {\bibinfo {volume} {3}},\ \bibinfo {pages} {625} (\bibinfo {year} {2021})}\BibitemShut {NoStop}%
\bibitem [{\citenamefont {Katabarwa}\ \emph {et~al.}(2024)\citenamefont {Katabarwa}, \citenamefont {Gratsea}, \citenamefont {Caesura},\ and\ \citenamefont {Johnson}}]{katabarwa_early_2024}%
  \BibitemOpen
  \bibfield  {author} {\bibinfo {author} {\bibfnamefont {A.}~\bibnamefont {Katabarwa}}, \bibinfo {author} {\bibfnamefont {K.}~\bibnamefont {Gratsea}}, \bibinfo {author} {\bibfnamefont {A.}~\bibnamefont {Caesura}},\ and\ \bibinfo {author} {\bibfnamefont {P.~D.}\ \bibnamefont {Johnson}},\ }\bibfield  {title} {\bibinfo {title} {Early {{Fault-Tolerant Quantum Computing}}},\ }\href {https://doi.org/10.1103/PRXQuantum.5.020101} {\bibfield  {journal} {\bibinfo  {journal} {PRX Quantum}\ }\textbf {\bibinfo {volume} {5}},\ \bibinfo {pages} {020101} (\bibinfo {year} {2024})}\BibitemShut {NoStop}%
\bibitem [{\citenamefont {Dob{\v s}{\'i}{\v c}ek}\ \emph {et~al.}(2007)\citenamefont {Dob{\v s}{\'i}{\v c}ek}, \citenamefont {Johansson}, \citenamefont {Shumeiko},\ and\ \citenamefont {Wendin}}]{dobsicek_arbitrary_2007}%
  \BibitemOpen
  \bibfield  {author} {\bibinfo {author} {\bibfnamefont {M.}~\bibnamefont {Dob{\v s}{\'i}{\v c}ek}}, \bibinfo {author} {\bibfnamefont {G.}~\bibnamefont {Johansson}}, \bibinfo {author} {\bibfnamefont {V.}~\bibnamefont {Shumeiko}},\ and\ \bibinfo {author} {\bibfnamefont {G.}~\bibnamefont {Wendin}},\ }\bibfield  {title} {\bibinfo {title} {Arbitrary accuracy iterative quantum phase estimation algorithm using a single ancillary qubit: {{A}} two-qubit benchmark},\ }\href {https://doi.org/10.1103/PhysRevA.76.030306} {\bibfield  {journal} {\bibinfo  {journal} {Phys. Rev. A}\ }\textbf {\bibinfo {volume} {76}},\ \bibinfo {pages} {030306} (\bibinfo {year} {2007})}\BibitemShut {NoStop}%
\bibitem [{\citenamefont {Svore}\ \emph {et~al.}(2013)\citenamefont {Svore}, \citenamefont {Hastings},\ and\ \citenamefont {Freedman}}]{svore_faster_2013}%
  \BibitemOpen
  \bibfield  {author} {\bibinfo {author} {\bibfnamefont {K.~M.}\ \bibnamefont {Svore}}, \bibinfo {author} {\bibfnamefont {M.~B.}\ \bibnamefont {Hastings}},\ and\ \bibinfo {author} {\bibfnamefont {M.}~\bibnamefont {Freedman}},\ }\bibfield  {title} {\bibinfo {title} {Faster {{Phase Estimation}}},\ }\bibfield  {journal} {\bibinfo  {journal} {arXiv preprint}\ }\href {https://doi.org/10.48550/arXiv.1304.0741} {10.48550/arXiv.1304.0741} (\bibinfo {year} {2013}),\ \Eprint {https://arxiv.org/abs/1304.0741} {arXiv:1304.0741} \BibitemShut {NoStop}%
\bibitem [{\citenamefont {Ding}\ and\ \citenamefont {Lin}(2023)}]{ding_even_2023}%
  \BibitemOpen
  \bibfield  {author} {\bibinfo {author} {\bibfnamefont {Z.}~\bibnamefont {Ding}}\ and\ \bibinfo {author} {\bibfnamefont {L.}~\bibnamefont {Lin}},\ }\bibfield  {title} {\bibinfo {title} {Even {{Shorter Quantum Circuit}} for {{Phase Estimation}} on {{Early Fault-Tolerant Quantum Computers}} with {{Applications}} to {{Ground-State Energy Estimation}}},\ }\href {https://doi.org/10.1103/PRXQuantum.4.020331} {\bibfield  {journal} {\bibinfo  {journal} {PRX Quantum}\ }\textbf {\bibinfo {volume} {4}},\ \bibinfo {pages} {020331} (\bibinfo {year} {2023})}\BibitemShut {NoStop}%
\bibitem [{\citenamefont {Ni}\ \emph {et~al.}(2023)\citenamefont {Ni}, \citenamefont {Li},\ and\ \citenamefont {Ying}}]{ni_lowdepth_2023}%
  \BibitemOpen
  \bibfield  {author} {\bibinfo {author} {\bibfnamefont {H.}~\bibnamefont {Ni}}, \bibinfo {author} {\bibfnamefont {H.}~\bibnamefont {Li}},\ and\ \bibinfo {author} {\bibfnamefont {L.}~\bibnamefont {Ying}},\ }\bibfield  {title} {\bibinfo {title} {On low-depth algorithms for quantum phase estimation},\ }\href {https://doi.org/10.22331/q-2023-11-06-1165} {\bibfield  {journal} {\bibinfo  {journal} {Quantum}\ }\textbf {\bibinfo {volume} {7}},\ \bibinfo {pages} {1165} (\bibinfo {year} {2023})}\BibitemShut {NoStop}%
\bibitem [{\citenamefont {Liu}\ \emph {et~al.}(2021)\citenamefont {Liu}, \citenamefont {Arunachalam},\ and\ \citenamefont {Temme}}]{liu_rigorous_2021}%
  \BibitemOpen
  \bibfield  {author} {\bibinfo {author} {\bibfnamefont {Y.}~\bibnamefont {Liu}}, \bibinfo {author} {\bibfnamefont {S.}~\bibnamefont {Arunachalam}},\ and\ \bibinfo {author} {\bibfnamefont {K.}~\bibnamefont {Temme}},\ }\bibfield  {title} {\bibinfo {title} {A rigorous and robust quantum speed-up in supervised machine learning},\ }\href {https://doi.org/10.1038/s41567-021-01287-z} {\bibfield  {journal} {\bibinfo  {journal} {Nat. Phys.}\ }\textbf {\bibinfo {volume} {17}},\ \bibinfo {pages} {1013} (\bibinfo {year} {2021})}\BibitemShut {NoStop}%
\bibitem [{\citenamefont {Paine}\ \emph {et~al.}(2023)\citenamefont {Paine}, \citenamefont {Elfving},\ and\ \citenamefont {Kyriienko}}]{paine_quantum_2023}%
  \BibitemOpen
  \bibfield  {author} {\bibinfo {author} {\bibfnamefont {A.~E.}\ \bibnamefont {Paine}}, \bibinfo {author} {\bibfnamefont {V.~E.}\ \bibnamefont {Elfving}},\ and\ \bibinfo {author} {\bibfnamefont {O.}~\bibnamefont {Kyriienko}},\ }\bibfield  {title} {\bibinfo {title} {Quantum kernel methods for solving regression problems and differential equations},\ }\href {https://doi.org/10.1103/PhysRevA.107.032428} {\bibfield  {journal} {\bibinfo  {journal} {Phys. Rev. A}\ }\textbf {\bibinfo {volume} {107}},\ \bibinfo {pages} {032428} (\bibinfo {year} {2023})}\BibitemShut {NoStop}%
\bibitem [{\citenamefont {Wierichs}\ \emph {et~al.}(2022)\citenamefont {Wierichs}, \citenamefont {Izaac}, \citenamefont {Wang},\ and\ \citenamefont {Lin}}]{wierichs_general_2022}%
  \BibitemOpen
  \bibfield  {author} {\bibinfo {author} {\bibfnamefont {D.}~\bibnamefont {Wierichs}}, \bibinfo {author} {\bibfnamefont {J.}~\bibnamefont {Izaac}}, \bibinfo {author} {\bibfnamefont {C.}~\bibnamefont {Wang}},\ and\ \bibinfo {author} {\bibfnamefont {C.~Y.-Y.}\ \bibnamefont {Lin}},\ }\bibfield  {title} {\bibinfo {title} {General parameter-shift rules for quantum gradients},\ }\href {https://doi.org/10.22331/q-2022-03-30-677} {\bibfield  {journal} {\bibinfo  {journal} {Quantum}\ }\textbf {\bibinfo {volume} {6}},\ \bibinfo {pages} {677} (\bibinfo {year} {2022})}\BibitemShut {NoStop}%
\bibitem [{\citenamefont {Li}\ \emph {et~al.}(2024)\citenamefont {Li}, \citenamefont {Dulal}, \citenamefont {Ohorodnikov}, \citenamefont {Wang},\ and\ \citenamefont {Ding}}]{li_efficient_2024}%
  \BibitemOpen
  \bibfield  {author} {\bibinfo {author} {\bibfnamefont {D.}~\bibnamefont {Li}}, \bibinfo {author} {\bibfnamefont {D.}~\bibnamefont {Dulal}}, \bibinfo {author} {\bibfnamefont {M.}~\bibnamefont {Ohorodnikov}}, \bibinfo {author} {\bibfnamefont {H.}~\bibnamefont {Wang}},\ and\ \bibinfo {author} {\bibfnamefont {Y.}~\bibnamefont {Ding}},\ }\bibfield  {title} {\bibinfo {title} {Efficient {{Quantum Gradient}} and {{Higher-order Derivative Estimation}} via {{Generalized Hadamard Test}}},\ }\bibfield  {journal} {\bibinfo  {journal} {arXiv preprint}\ }\href {https://doi.org/10.48550/arXiv.2408.05406} {10.48550/arXiv.2408.05406} (\bibinfo {year} {2024}),\ \Eprint {https://arxiv.org/abs/2408.05406} {arXiv:2408.05406} \BibitemShut {NoStop}%
\bibitem [{\citenamefont {Verteletskyi}\ \emph {et~al.}(2020)\citenamefont {Verteletskyi}, \citenamefont {Yen},\ and\ \citenamefont {Izmaylov}}]{verteletskyi_measurement_2020}%
  \BibitemOpen
  \bibfield  {author} {\bibinfo {author} {\bibfnamefont {V.}~\bibnamefont {Verteletskyi}}, \bibinfo {author} {\bibfnamefont {T.-C.}\ \bibnamefont {Yen}},\ and\ \bibinfo {author} {\bibfnamefont {A.~F.}\ \bibnamefont {Izmaylov}},\ }\bibfield  {title} {\bibinfo {title} {Measurement optimization in the variational quantum eigensolver using a minimum clique cover},\ }\href {https://doi.org/10.1063/1.5141458} {\bibfield  {journal} {\bibinfo  {journal} {J. Chem. Phys.}\ }\textbf {\bibinfo {volume} {152}},\ \bibinfo {pages} {124114} (\bibinfo {year} {2020})}\BibitemShut {NoStop}%
\bibitem [{\citenamefont {Heidari}\ \emph {et~al.}(2024)\citenamefont {Heidari}, \citenamefont {Mozakka},\ and\ \citenamefont {Szpankowski}}]{heidari_efficient_2024}%
  \BibitemOpen
  \bibfield  {author} {\bibinfo {author} {\bibfnamefont {M.}~\bibnamefont {Heidari}}, \bibinfo {author} {\bibfnamefont {M.}~\bibnamefont {Mozakka}},\ and\ \bibinfo {author} {\bibfnamefont {W.}~\bibnamefont {Szpankowski}},\ }\bibfield  {title} {\bibinfo {title} {Efficient {{Gradient Estimation}} of {{Variational Quantum Circuits}} with {{Lie Algebraic Symmetries}}},\ }\bibfield  {journal} {\bibinfo  {journal} {arXiv preprint}\ }\href {https://doi.org/10.48550/arXiv.2404.05108} {10.48550/arXiv.2404.05108} (\bibinfo {year} {2024}),\ \Eprint {https://arxiv.org/abs/2404.05108} {arXiv:2404.05108} \BibitemShut {NoStop}%
\bibitem [{\citenamefont {Peruzzo}\ \emph {et~al.}(2014)\citenamefont {Peruzzo}, \citenamefont {McClean}, \citenamefont {Shadbolt}, \citenamefont {Yung}, \citenamefont {Zhou}, \citenamefont {Love}, \citenamefont {{Aspuru-Guzik}},\ and\ \citenamefont {O'Brien}}]{peruzzo_variational_2014}%
  \BibitemOpen
  \bibfield  {author} {\bibinfo {author} {\bibfnamefont {A.}~\bibnamefont {Peruzzo}}, \bibinfo {author} {\bibfnamefont {J.}~\bibnamefont {McClean}}, \bibinfo {author} {\bibfnamefont {P.}~\bibnamefont {Shadbolt}}, \bibinfo {author} {\bibfnamefont {M.-H.}\ \bibnamefont {Yung}}, \bibinfo {author} {\bibfnamefont {X.-Q.}\ \bibnamefont {Zhou}}, \bibinfo {author} {\bibfnamefont {P.~J.}\ \bibnamefont {Love}}, \bibinfo {author} {\bibfnamefont {A.}~\bibnamefont {{Aspuru-Guzik}}},\ and\ \bibinfo {author} {\bibfnamefont {J.~L.}\ \bibnamefont {O'Brien}},\ }\bibfield  {title} {\bibinfo {title} {A variational eigenvalue solver on a photonic quantum processor},\ }\href {https://doi.org/10.1038/ncomms5213} {\bibfield  {journal} {\bibinfo  {journal} {Nat. Commun.}\ }\textbf {\bibinfo {volume} {5}},\ \bibinfo {pages} {4213} (\bibinfo {year} {2014})}\BibitemShut {NoStop}%
\bibitem [{\citenamefont {Tilly}\ \emph {et~al.}(2022)\citenamefont {Tilly}, \citenamefont {Chen}, \citenamefont {Cao}, \citenamefont {Picozzi}, \citenamefont {Setia}, \citenamefont {Li}, \citenamefont {Grant}, \citenamefont {Wossnig}, \citenamefont {Rungger}, \citenamefont {Booth},\ and\ \citenamefont {Tennyson}}]{tilly_variational_2022}%
  \BibitemOpen
  \bibfield  {author} {\bibinfo {author} {\bibfnamefont {J.}~\bibnamefont {Tilly}}, \bibinfo {author} {\bibfnamefont {H.}~\bibnamefont {Chen}}, \bibinfo {author} {\bibfnamefont {S.}~\bibnamefont {Cao}}, \bibinfo {author} {\bibfnamefont {D.}~\bibnamefont {Picozzi}}, \bibinfo {author} {\bibfnamefont {K.}~\bibnamefont {Setia}}, \bibinfo {author} {\bibfnamefont {Y.}~\bibnamefont {Li}}, \bibinfo {author} {\bibfnamefont {E.}~\bibnamefont {Grant}}, \bibinfo {author} {\bibfnamefont {L.}~\bibnamefont {Wossnig}}, \bibinfo {author} {\bibfnamefont {I.}~\bibnamefont {Rungger}}, \bibinfo {author} {\bibfnamefont {G.~H.}\ \bibnamefont {Booth}},\ and\ \bibinfo {author} {\bibfnamefont {J.}~\bibnamefont {Tennyson}},\ }\bibfield  {title} {\bibinfo {title} {The {{Variational Quantum Eigensolver}}: {{A}} review of methods and best practices},\ }\href {https://doi.org/10.1016/j.physrep.2022.08.003} {\bibfield  {journal} {\bibinfo  {journal} {Physics Reports}\ }\textbf {\bibinfo {volume} {986}},\ \bibinfo {pages} {1} (\bibinfo {year} {2022})}\BibitemShut {NoStop}%
\bibitem [{\citenamefont {Nelder}\ and\ \citenamefont {Mead}(1965)}]{nelder_simplex_1965}%
  \BibitemOpen
  \bibfield  {author} {\bibinfo {author} {\bibfnamefont {J.~A.}\ \bibnamefont {Nelder}}\ and\ \bibinfo {author} {\bibfnamefont {R.}~\bibnamefont {Mead}},\ }\bibfield  {title} {\bibinfo {title} {A {{Simplex Method}} for {{Function Minimization}}},\ }\href {https://doi.org/10.1093/comjnl/7.4.308} {\bibfield  {journal} {\bibinfo  {journal} {The Computer Journal}\ }\textbf {\bibinfo {volume} {7}},\ \bibinfo {pages} {308} (\bibinfo {year} {1965})}\BibitemShut {NoStop}%
\bibitem [{\citenamefont {Nakanishi}\ \emph {et~al.}(2020)\citenamefont {Nakanishi}, \citenamefont {Fujii},\ and\ \citenamefont {Todo}}]{nakanishi_sequential_2020}%
  \BibitemOpen
  \bibfield  {author} {\bibinfo {author} {\bibfnamefont {K.~M.}\ \bibnamefont {Nakanishi}}, \bibinfo {author} {\bibfnamefont {K.}~\bibnamefont {Fujii}},\ and\ \bibinfo {author} {\bibfnamefont {S.}~\bibnamefont {Todo}},\ }\bibfield  {title} {\bibinfo {title} {Sequential minimal optimization for quantum-classical hybrid algorithms},\ }\href {https://doi.org/10.1103/PhysRevResearch.2.043158} {\bibfield  {journal} {\bibinfo  {journal} {Phys. Rev. Res.}\ }\textbf {\bibinfo {volume} {2}},\ \bibinfo {pages} {043158} (\bibinfo {year} {2020})}\BibitemShut {NoStop}%
\bibitem [{\citenamefont {Ostaszewski}\ \emph {et~al.}(2021)\citenamefont {Ostaszewski}, \citenamefont {Grant},\ and\ \citenamefont {Benedetti}}]{ostaszewski_structure_2021}%
  \BibitemOpen
  \bibfield  {author} {\bibinfo {author} {\bibfnamefont {M.}~\bibnamefont {Ostaszewski}}, \bibinfo {author} {\bibfnamefont {E.}~\bibnamefont {Grant}},\ and\ \bibinfo {author} {\bibfnamefont {M.}~\bibnamefont {Benedetti}},\ }\bibfield  {title} {\bibinfo {title} {Structure optimization for parameterized quantum circuits},\ }\href {https://doi.org/10.22331/q-2021-01-28-391} {\bibfield  {journal} {\bibinfo  {journal} {Quantum}\ }\textbf {\bibinfo {volume} {5}},\ \bibinfo {pages} {391} (\bibinfo {year} {2021})}\BibitemShut {NoStop}%
\bibitem [{\citenamefont {Watanabe}\ \emph {et~al.}(2023)\citenamefont {Watanabe}, \citenamefont {Raymond}, \citenamefont {Ohnishi}, \citenamefont {Kaminishi},\ and\ \citenamefont {Sugawara}}]{watanabe_optimizing_2023}%
  \BibitemOpen
  \bibfield  {author} {\bibinfo {author} {\bibfnamefont {H.~C.}\ \bibnamefont {Watanabe}}, \bibinfo {author} {\bibfnamefont {R.}~\bibnamefont {Raymond}}, \bibinfo {author} {\bibfnamefont {Y.-y.}\ \bibnamefont {Ohnishi}}, \bibinfo {author} {\bibfnamefont {E.}~\bibnamefont {Kaminishi}},\ and\ \bibinfo {author} {\bibfnamefont {M.}~\bibnamefont {Sugawara}},\ }\bibfield  {title} {\bibinfo {title} {Optimizing {{Parameterized Quantum Circuits}} with {{Free-Axis Selection}}},\ }\bibfield  {journal} {\bibinfo  {journal} {arXiv preprint}\ }\href {https://doi.org/10.48550/arXiv.2104.14875} {10.48550/arXiv.2104.14875} (\bibinfo {year} {2023}),\ \Eprint {https://arxiv.org/abs/2104.14875} {arXiv:2104.14875} \BibitemShut {NoStop}%
\bibitem [{\citenamefont {Wada}\ \emph {et~al.}(2022)\citenamefont {Wada}, \citenamefont {Raymond}, \citenamefont {Ohnishi}, \citenamefont {Kaminishi}, \citenamefont {Sugawara}, \citenamefont {Yamamoto},\ and\ \citenamefont {Watanabe}}]{wada_simulating_2022}%
  \BibitemOpen
  \bibfield  {author} {\bibinfo {author} {\bibfnamefont {K.}~\bibnamefont {Wada}}, \bibinfo {author} {\bibfnamefont {R.}~\bibnamefont {Raymond}}, \bibinfo {author} {\bibfnamefont {Y.-y.}\ \bibnamefont {Ohnishi}}, \bibinfo {author} {\bibfnamefont {E.}~\bibnamefont {Kaminishi}}, \bibinfo {author} {\bibfnamefont {M.}~\bibnamefont {Sugawara}}, \bibinfo {author} {\bibfnamefont {N.}~\bibnamefont {Yamamoto}},\ and\ \bibinfo {author} {\bibfnamefont {H.~C.}\ \bibnamefont {Watanabe}},\ }\bibfield  {title} {\bibinfo {title} {Simulating time evolution with fully optimized single-qubit gates on parametrized quantum circuits},\ }\href {https://doi.org/10.1103/PhysRevA.105.062421} {\bibfield  {journal} {\bibinfo  {journal} {Phys. Rev. A}\ }\textbf {\bibinfo {volume} {105}},\ \bibinfo {pages} {062421} (\bibinfo {year} {2022})}\BibitemShut {NoStop}%
\bibitem [{\citenamefont {Parrish}\ \emph {et~al.}(2019)\citenamefont {Parrish}, \citenamefont {Iosue}, \citenamefont {Ozaeta},\ and\ \citenamefont {McMahon}}]{parrish_jacobi_2019}%
  \BibitemOpen
  \bibfield  {author} {\bibinfo {author} {\bibfnamefont {R.~M.}\ \bibnamefont {Parrish}}, \bibinfo {author} {\bibfnamefont {J.~T.}\ \bibnamefont {Iosue}}, \bibinfo {author} {\bibfnamefont {A.}~\bibnamefont {Ozaeta}},\ and\ \bibinfo {author} {\bibfnamefont {P.~L.}\ \bibnamefont {McMahon}},\ }\bibfield  {title} {\bibinfo {title} {A {{Jacobi Diagonalization}} and {{Anderson Acceleration Algorithm For Variational Quantum Algorithm Parameter Optimization}}},\ }\bibfield  {journal} {\bibinfo  {journal} {arXiv preprint}\ }\href {https://doi.org/10.48550/arXiv.1904.03206} {10.48550/arXiv.1904.03206} (\bibinfo {year} {2019}),\ \Eprint {https://arxiv.org/abs/1904.03206} {arXiv:1904.03206} \BibitemShut {NoStop}%
\bibitem [{\citenamefont {Harrow}\ and\ \citenamefont {Napp}(2021)}]{harrow_lowdepth_2021}%
  \BibitemOpen
  \bibfield  {author} {\bibinfo {author} {\bibfnamefont {A.~W.}\ \bibnamefont {Harrow}}\ and\ \bibinfo {author} {\bibfnamefont {J.~C.}\ \bibnamefont {Napp}},\ }\bibfield  {title} {\bibinfo {title} {Low-{{Depth Gradient Measurements Can Improve Convergence}} in {{Variational Hybrid Quantum-Classical Algorithms}}},\ }\href {https://doi.org/10.1103/PhysRevLett.126.140502} {\bibfield  {journal} {\bibinfo  {journal} {Phys. Rev. Lett.}\ }\textbf {\bibinfo {volume} {126}},\ \bibinfo {pages} {140502} (\bibinfo {year} {2021})}\BibitemShut {NoStop}%
\bibitem [{\citenamefont {Guerreschi}\ and\ \citenamefont {Smelyanskiy}(2017)}]{guerreschi_practical_2017}%
  \BibitemOpen
  \bibfield  {author} {\bibinfo {author} {\bibfnamefont {G.~G.}\ \bibnamefont {Guerreschi}}\ and\ \bibinfo {author} {\bibfnamefont {M.}~\bibnamefont {Smelyanskiy}},\ }\bibfield  {title} {\bibinfo {title} {Practical optimization for hybrid quantum-classical algorithms},\ }\bibfield  {journal} {\bibinfo  {journal} {arXiv preprint}\ }\href {https://doi.org/10.48550/arXiv.1701.01450} {10.48550/arXiv.1701.01450} (\bibinfo {year} {2017}),\ \Eprint {https://arxiv.org/abs/1701.01450} {arXiv:1701.01450} \BibitemShut {NoStop}%
\bibitem [{\citenamefont {Romero}\ \emph {et~al.}(2018)\citenamefont {Romero}, \citenamefont {Babbush}, \citenamefont {McClean}, \citenamefont {Hempel}, \citenamefont {Love},\ and\ \citenamefont {{Aspuru-Guzik}}}]{romero_strategies_2018}%
  \BibitemOpen
  \bibfield  {author} {\bibinfo {author} {\bibfnamefont {J.}~\bibnamefont {Romero}}, \bibinfo {author} {\bibfnamefont {R.}~\bibnamefont {Babbush}}, \bibinfo {author} {\bibfnamefont {J.~R.}\ \bibnamefont {McClean}}, \bibinfo {author} {\bibfnamefont {C.}~\bibnamefont {Hempel}}, \bibinfo {author} {\bibfnamefont {P.~J.}\ \bibnamefont {Love}},\ and\ \bibinfo {author} {\bibfnamefont {A.}~\bibnamefont {{Aspuru-Guzik}}},\ }\bibfield  {title} {\bibinfo {title} {Strategies for quantum computing molecular energies using the unitary coupled cluster ansatz},\ }\href {https://doi.org/10.1088/2058-9565/aad3e4} {\bibfield  {journal} {\bibinfo  {journal} {Quantum Sci. Technol.}\ }\textbf {\bibinfo {volume} {4}},\ \bibinfo {pages} {014008} (\bibinfo {year} {2018})}\BibitemShut {NoStop}%
\bibitem [{\citenamefont {Schuld}\ \emph {et~al.}(2019)\citenamefont {Schuld}, \citenamefont {Bergholm}, \citenamefont {Gogolin}, \citenamefont {Izaac},\ and\ \citenamefont {Killoran}}]{schuld_evaluating_2019}%
  \BibitemOpen
  \bibfield  {author} {\bibinfo {author} {\bibfnamefont {M.}~\bibnamefont {Schuld}}, \bibinfo {author} {\bibfnamefont {V.}~\bibnamefont {Bergholm}}, \bibinfo {author} {\bibfnamefont {C.}~\bibnamefont {Gogolin}}, \bibinfo {author} {\bibfnamefont {J.}~\bibnamefont {Izaac}},\ and\ \bibinfo {author} {\bibfnamefont {N.}~\bibnamefont {Killoran}},\ }\bibfield  {title} {\bibinfo {title} {Evaluating analytic gradients on quantum hardware},\ }\href {https://doi.org/10.1103/PhysRevA.99.032331} {\bibfield  {journal} {\bibinfo  {journal} {Phys. Rev. A}\ }\textbf {\bibinfo {volume} {99}},\ \bibinfo {pages} {032331} (\bibinfo {year} {2019})}\BibitemShut {NoStop}%
\bibitem [{\citenamefont {Mitarai}\ \emph {et~al.}(2018)\citenamefont {Mitarai}, \citenamefont {Negoro}, \citenamefont {Kitagawa},\ and\ \citenamefont {Fujii}}]{mitarai_quantum_2018}%
  \BibitemOpen
  \bibfield  {author} {\bibinfo {author} {\bibfnamefont {K.}~\bibnamefont {Mitarai}}, \bibinfo {author} {\bibfnamefont {M.}~\bibnamefont {Negoro}}, \bibinfo {author} {\bibfnamefont {M.}~\bibnamefont {Kitagawa}},\ and\ \bibinfo {author} {\bibfnamefont {K.}~\bibnamefont {Fujii}},\ }\bibfield  {title} {\bibinfo {title} {Quantum circuit learning},\ }\href {https://doi.org/10.1103/PhysRevA.98.032309} {\bibfield  {journal} {\bibinfo  {journal} {Phys. Rev. A}\ }\textbf {\bibinfo {volume} {98}},\ \bibinfo {pages} {032309} (\bibinfo {year} {2018})}\BibitemShut {NoStop}%
\bibitem [{\citenamefont {Hubregtsen}\ \emph {et~al.}(2022)\citenamefont {Hubregtsen}, \citenamefont {Wilde}, \citenamefont {Qasim},\ and\ \citenamefont {Eisert}}]{hubregtsen_singlecomponent_2022}%
  \BibitemOpen
  \bibfield  {author} {\bibinfo {author} {\bibfnamefont {T.}~\bibnamefont {Hubregtsen}}, \bibinfo {author} {\bibfnamefont {F.}~\bibnamefont {Wilde}}, \bibinfo {author} {\bibfnamefont {S.}~\bibnamefont {Qasim}},\ and\ \bibinfo {author} {\bibfnamefont {J.}~\bibnamefont {Eisert}},\ }\bibfield  {title} {\bibinfo {title} {Single-component gradient rules for variational quantum algorithms},\ }\href {https://doi.org/10.1088/2058-9565/ac6824} {\bibfield  {journal} {\bibinfo  {journal} {Quantum Sci. Technol.}\ }\textbf {\bibinfo {volume} {7}},\ \bibinfo {pages} {035008} (\bibinfo {year} {2022})}\BibitemShut {NoStop}%
\bibitem [{\citenamefont {Crooks}(2019)}]{crooks_gradients_2019}%
  \BibitemOpen
  \bibfield  {author} {\bibinfo {author} {\bibfnamefont {G.~E.}\ \bibnamefont {Crooks}},\ }\bibfield  {title} {\bibinfo {title} {Gradients of parameterized quantum gates using the parameter-shift rule and gate decomposition},\ }\bibfield  {journal} {\bibinfo  {journal} {arXiv preprint}\ }\href {https://doi.org/10.48550/arXiv.1905.13311} {10.48550/arXiv.1905.13311} (\bibinfo {year} {2019}),\ \Eprint {https://arxiv.org/abs/1905.13311} {arXiv:1905.13311} \BibitemShut {NoStop}%
\bibitem [{\citenamefont {Izmaylov}\ \emph {et~al.}(2021)\citenamefont {Izmaylov}, \citenamefont {Lang},\ and\ \citenamefont {Yen}}]{izmaylov_analytic_2021}%
  \BibitemOpen
  \bibfield  {author} {\bibinfo {author} {\bibfnamefont {A.~F.}\ \bibnamefont {Izmaylov}}, \bibinfo {author} {\bibfnamefont {R.~A.}\ \bibnamefont {Lang}},\ and\ \bibinfo {author} {\bibfnamefont {T.-C.}\ \bibnamefont {Yen}},\ }\bibfield  {title} {\bibinfo {title} {Analytic gradients in variational quantum algorithms: {{Algebraic}} extensions of the parameter-shift rule to general unitary transformations},\ }\href {https://doi.org/10.1103/PhysRevA.104.062443} {\bibfield  {journal} {\bibinfo  {journal} {Phys. Rev. A}\ }\textbf {\bibinfo {volume} {104}},\ \bibinfo {pages} {062443} (\bibinfo {year} {2021})}\BibitemShut {NoStop}%
\bibitem [{\citenamefont {Nielsen}\ and\ \citenamefont {Chuang}(2010)}]{nielsen_quantum_2010}%
  \BibitemOpen
  \bibfield  {author} {\bibinfo {author} {\bibfnamefont {M.~A.}\ \bibnamefont {Nielsen}}\ and\ \bibinfo {author} {\bibfnamefont {I.~L.}\ \bibnamefont {Chuang}},\ }\href {https://doi.org/10.1017/CBO9780511976667} {\emph {\bibinfo {title} {Quantum {{Computation}} and {{Quantum Information}}}}}\ (\bibinfo  {publisher} {Cambridge University Press},\ \bibinfo {year} {2010})\BibitemShut {NoStop}%
\bibitem [{\citenamefont {Kshirsagar}\ \emph {et~al.}(2024)\citenamefont {Kshirsagar}, \citenamefont {Katabarwa},\ and\ \citenamefont {Johnson}}]{kshirsagar_proving_2024}%
  \BibitemOpen
  \bibfield  {author} {\bibinfo {author} {\bibfnamefont {R.}~\bibnamefont {Kshirsagar}}, \bibinfo {author} {\bibfnamefont {A.}~\bibnamefont {Katabarwa}},\ and\ \bibinfo {author} {\bibfnamefont {P.~D.}\ \bibnamefont {Johnson}},\ }\bibfield  {title} {\bibinfo {title} {On proving the robustness of algorithms for early fault-tolerant quantum computers},\ }\href {https://doi.org/10.22331/q-2024-11-20-1531} {\bibfield  {journal} {\bibinfo  {journal} {Quantum}\ }\textbf {\bibinfo {volume} {8}},\ \bibinfo {pages} {1531} (\bibinfo {year} {2024})}\BibitemShut {NoStop}%
\bibitem [{\citenamefont {Lin}\ and\ \citenamefont {Tong}(2022)}]{lin_heisenberglimited_2022}%
  \BibitemOpen
  \bibfield  {author} {\bibinfo {author} {\bibfnamefont {L.}~\bibnamefont {Lin}}\ and\ \bibinfo {author} {\bibfnamefont {Y.}~\bibnamefont {Tong}},\ }\bibfield  {title} {\bibinfo {title} {Heisenberg-{{Limited Ground-State Energy Estimation}} for {{Early Fault-Tolerant Quantum Computers}}},\ }\href {https://doi.org/10.1103/PRXQuantum.3.010318} {\bibfield  {journal} {\bibinfo  {journal} {PRX Quantum}\ }\textbf {\bibinfo {volume} {3}},\ \bibinfo {pages} {010318} (\bibinfo {year} {2022})}\BibitemShut {NoStop}%
\bibitem [{\citenamefont {Wan}\ \emph {et~al.}(2022)\citenamefont {Wan}, \citenamefont {Berta},\ and\ \citenamefont {Campbell}}]{wan_randomized_2022}%
  \BibitemOpen
  \bibfield  {author} {\bibinfo {author} {\bibfnamefont {K.}~\bibnamefont {Wan}}, \bibinfo {author} {\bibfnamefont {M.}~\bibnamefont {Berta}},\ and\ \bibinfo {author} {\bibfnamefont {E.~T.}\ \bibnamefont {Campbell}},\ }\bibfield  {title} {\bibinfo {title} {Randomized {{Quantum Algorithm}} for {{Statistical Phase Estimation}}},\ }\href {https://doi.org/10.1103/PhysRevLett.129.030503} {\bibfield  {journal} {\bibinfo  {journal} {Phys. Rev. Lett.}\ }\textbf {\bibinfo {volume} {129}},\ \bibinfo {pages} {030503} (\bibinfo {year} {2022})}\BibitemShut {NoStop}%
\bibitem [{\citenamefont {Dutkiewicz}\ \emph {et~al.}(2022)\citenamefont {Dutkiewicz}, \citenamefont {Terhal},\ and\ \citenamefont {O'Brien}}]{dutkiewicz_heisenberglimited_2022}%
  \BibitemOpen
  \bibfield  {author} {\bibinfo {author} {\bibfnamefont {A.}~\bibnamefont {Dutkiewicz}}, \bibinfo {author} {\bibfnamefont {B.~M.}\ \bibnamefont {Terhal}},\ and\ \bibinfo {author} {\bibfnamefont {T.~E.}\ \bibnamefont {O'Brien}},\ }\bibfield  {title} {\bibinfo {title} {Heisenberg-limited quantum phase estimation of multiple eigenvalues with few control qubits},\ }\href {https://doi.org/10.22331/q-2022-10-06-830} {\bibfield  {journal} {\bibinfo  {journal} {Quantum}\ }\textbf {\bibinfo {volume} {6}},\ \bibinfo {pages} {830} (\bibinfo {year} {2022})}\BibitemShut {NoStop}%
\bibitem [{\citenamefont {O'Brien}\ \emph {et~al.}(2019)\citenamefont {O'Brien}, \citenamefont {Tarasinski},\ and\ \citenamefont {Terhal}}]{obrien_quantum_2019}%
  \BibitemOpen
  \bibfield  {author} {\bibinfo {author} {\bibfnamefont {T.~E.}\ \bibnamefont {O'Brien}}, \bibinfo {author} {\bibfnamefont {B.}~\bibnamefont {Tarasinski}},\ and\ \bibinfo {author} {\bibfnamefont {B.~M.}\ \bibnamefont {Terhal}},\ }\bibfield  {title} {\bibinfo {title} {Quantum phase estimation of multiple eigenvalues for small-scale (noisy) experiments},\ }\href {https://doi.org/10.1088/1367-2630/aafb8e} {\bibfield  {journal} {\bibinfo  {journal} {New J. Phys.}\ }\textbf {\bibinfo {volume} {21}},\ \bibinfo {pages} {023022} (\bibinfo {year} {2019})}\BibitemShut {NoStop}%
\bibitem [{\citenamefont {Toshio}\ \emph {et~al.}(2025)\citenamefont {Toshio}, \citenamefont {Akahoshi}, \citenamefont {Fujisaki}, \citenamefont {Oshima}, \citenamefont {Sato},\ and\ \citenamefont {Fujii}}]{toshio_practical_2025}%
  \BibitemOpen
  \bibfield  {author} {\bibinfo {author} {\bibfnamefont {R.}~\bibnamefont {Toshio}}, \bibinfo {author} {\bibfnamefont {Y.}~\bibnamefont {Akahoshi}}, \bibinfo {author} {\bibfnamefont {J.}~\bibnamefont {Fujisaki}}, \bibinfo {author} {\bibfnamefont {H.}~\bibnamefont {Oshima}}, \bibinfo {author} {\bibfnamefont {S.}~\bibnamefont {Sato}},\ and\ \bibinfo {author} {\bibfnamefont {K.}~\bibnamefont {Fujii}},\ }\bibfield  {title} {\bibinfo {title} {Practical {{Quantum Advantage}} on {{Partially Fault-Tolerant Quantum Computer}}},\ }\href {https://doi.org/10.1103/PhysRevX.15.021057} {\bibfield  {journal} {\bibinfo  {journal} {Phys. Rev. X}\ }\textbf {\bibinfo {volume} {15}},\ \bibinfo {pages} {021057} (\bibinfo {year} {2025})}\BibitemShut {NoStop}%
\bibitem [{\citenamefont {Nelson}\ and\ \citenamefont {Baczewski}(2024)}]{nelson_assessment_2024}%
  \BibitemOpen
  \bibfield  {author} {\bibinfo {author} {\bibfnamefont {J.~S.}\ \bibnamefont {Nelson}}\ and\ \bibinfo {author} {\bibfnamefont {A.~D.}\ \bibnamefont {Baczewski}},\ }\bibfield  {title} {\bibinfo {title} {Assessment of quantum phase estimation protocols for early fault-tolerant quantum computers},\ }\href {https://doi.org/10.1103/PhysRevA.110.042420} {\bibfield  {journal} {\bibinfo  {journal} {Phys. Rev. A}\ }\textbf {\bibinfo {volume} {110}},\ \bibinfo {pages} {042420} (\bibinfo {year} {2024})}\BibitemShut {NoStop}%
\bibitem [{\citenamefont {Liang}\ \emph {et~al.}(2024)\citenamefont {Liang}, \citenamefont {Zhou}, \citenamefont {Dalal},\ and\ \citenamefont {Johnson}}]{liang_modeling_2024}%
  \BibitemOpen
  \bibfield  {author} {\bibinfo {author} {\bibfnamefont {Q.}~\bibnamefont {Liang}}, \bibinfo {author} {\bibfnamefont {Y.}~\bibnamefont {Zhou}}, \bibinfo {author} {\bibfnamefont {A.}~\bibnamefont {Dalal}},\ and\ \bibinfo {author} {\bibfnamefont {P.}~\bibnamefont {Johnson}},\ }\bibfield  {title} {\bibinfo {title} {Modeling the performance of early fault-tolerant quantum algorithms},\ }\href {https://doi.org/10.1103/PhysRevResearch.6.023118} {\bibfield  {journal} {\bibinfo  {journal} {Phys. Rev. Res.}\ }\textbf {\bibinfo {volume} {6}},\ \bibinfo {pages} {023118} (\bibinfo {year} {2024})}\BibitemShut {NoStop}%
\bibitem [{\citenamefont {Ku}\ \emph {et~al.}(2025)\citenamefont {Ku}, \citenamefont {Chen}, \citenamefont {Hu},\ and\ \citenamefont {Hsieh}}]{ku_benchmarking_2025}%
  \BibitemOpen
  \bibfield  {author} {\bibinfo {author} {\bibfnamefont {C.}~\bibnamefont {Ku}}, \bibinfo {author} {\bibfnamefont {Y.-C.}\ \bibnamefont {Chen}}, \bibinfo {author} {\bibfnamefont {A.}~\bibnamefont {Hu}},\ and\ \bibinfo {author} {\bibfnamefont {M.-H.}\ \bibnamefont {Hsieh}},\ }\bibfield  {title} {\bibinfo {title} {Benchmarking {{Quantum Simulation Methods}}},\ }\bibfield  {journal} {\bibinfo  {journal} {arXiv preprint}\ }\href {https://doi.org/10.48550/arXiv.2510.01710} {10.48550/arXiv.2510.01710} (\bibinfo {year} {2025}),\ \Eprint {https://arxiv.org/abs/2510.01710} {arXiv:2510.01710} \BibitemShut {NoStop}%
\bibitem [{\citenamefont {Vapnik}(2000)}]{vapnik_nature_2000}%
  \BibitemOpen
  \bibfield  {author} {\bibinfo {author} {\bibfnamefont {V.~N.}\ \bibnamefont {Vapnik}},\ }\href {https://doi.org/10.1007/978-1-4757-3264-1} {\emph {\bibinfo {title} {The {{Nature}} of {{Statistical Learning Theory}}}}}\ (\bibinfo  {publisher} {Springer},\ \bibinfo {address} {New York, NY},\ \bibinfo {year} {2000})\BibitemShut {NoStop}%
\bibitem [{\citenamefont {Havl{\'i}{\v c}ek}\ \emph {et~al.}(2019)\citenamefont {Havl{\'i}{\v c}ek}, \citenamefont {C{\'o}rcoles}, \citenamefont {Temme}, \citenamefont {Harrow}, \citenamefont {Kandala}, \citenamefont {Chow},\ and\ \citenamefont {Gambetta}}]{havlicek_supervised_2019}%
  \BibitemOpen
  \bibfield  {author} {\bibinfo {author} {\bibfnamefont {V.}~\bibnamefont {Havl{\'i}{\v c}ek}}, \bibinfo {author} {\bibfnamefont {A.~D.}\ \bibnamefont {C{\'o}rcoles}}, \bibinfo {author} {\bibfnamefont {K.}~\bibnamefont {Temme}}, \bibinfo {author} {\bibfnamefont {A.~W.}\ \bibnamefont {Harrow}}, \bibinfo {author} {\bibfnamefont {A.}~\bibnamefont {Kandala}}, \bibinfo {author} {\bibfnamefont {J.~M.}\ \bibnamefont {Chow}},\ and\ \bibinfo {author} {\bibfnamefont {J.~M.}\ \bibnamefont {Gambetta}},\ }\bibfield  {title} {\bibinfo {title} {Supervised learning with quantum-enhanced feature spaces},\ }\href {https://doi.org/10.1038/s41586-019-0980-2} {\bibfield  {journal} {\bibinfo  {journal} {Nature}\ }\textbf {\bibinfo {volume} {567}},\ \bibinfo {pages} {209} (\bibinfo {year} {2019})}\BibitemShut {NoStop}%
\bibitem [{\citenamefont {D{\"u}r}\ \emph {et~al.}(2000)\citenamefont {D{\"u}r}, \citenamefont {Vidal},\ and\ \citenamefont {Cirac}}]{dur_three_2000}%
  \BibitemOpen
  \bibfield  {author} {\bibinfo {author} {\bibfnamefont {W.}~\bibnamefont {D{\"u}r}}, \bibinfo {author} {\bibfnamefont {G.}~\bibnamefont {Vidal}},\ and\ \bibinfo {author} {\bibfnamefont {J.~I.}\ \bibnamefont {Cirac}},\ }\bibfield  {title} {\bibinfo {title} {Three qubits can be entangled in two inequivalent ways},\ }\href {https://doi.org/10.1103/PhysRevA.62.062314} {\bibfield  {journal} {\bibinfo  {journal} {Phys. Rev. A}\ }\textbf {\bibinfo {volume} {62}},\ \bibinfo {pages} {062314} (\bibinfo {year} {2000})}\BibitemShut {NoStop}%
\bibitem [{\citenamefont {Cruz}\ \emph {et~al.}(2019)\citenamefont {Cruz}, \citenamefont {Fournier}, \citenamefont {Gremion}, \citenamefont {Jeannerot}, \citenamefont {Komagata}, \citenamefont {Tosic}, \citenamefont {Thiesbrummel}, \citenamefont {Chan}, \citenamefont {Macris}, \citenamefont {Dupertuis},\ and\ \citenamefont {{Javerzac-Galy}}}]{cruz_efficient_2019}%
  \BibitemOpen
  \bibfield  {author} {\bibinfo {author} {\bibfnamefont {D.}~\bibnamefont {Cruz}}, \bibinfo {author} {\bibfnamefont {R.}~\bibnamefont {Fournier}}, \bibinfo {author} {\bibfnamefont {F.}~\bibnamefont {Gremion}}, \bibinfo {author} {\bibfnamefont {A.}~\bibnamefont {Jeannerot}}, \bibinfo {author} {\bibfnamefont {K.}~\bibnamefont {Komagata}}, \bibinfo {author} {\bibfnamefont {T.}~\bibnamefont {Tosic}}, \bibinfo {author} {\bibfnamefont {J.}~\bibnamefont {Thiesbrummel}}, \bibinfo {author} {\bibfnamefont {C.~L.}\ \bibnamefont {Chan}}, \bibinfo {author} {\bibfnamefont {N.}~\bibnamefont {Macris}}, \bibinfo {author} {\bibfnamefont {M.-A.}\ \bibnamefont {Dupertuis}},\ and\ \bibinfo {author} {\bibfnamefont {C.}~\bibnamefont {{Javerzac-Galy}}},\ }\bibfield  {title} {\bibinfo {title} {Efficient {{Quantum Algorithms}} for {{GHZ}} and {{W States}}, and {{Implementation}} on the {{IBM Quantum Computer}}},\ }\href {https://doi.org/10.1002/qute.201900015} {\bibfield  {journal} {\bibinfo  {journal} {Advanced Quantum Technologies}\ }\textbf {\bibinfo {volume} {2}},\ \bibinfo {pages} {1900015} (\bibinfo {year} {2019})}\BibitemShut {NoStop}%
\bibitem [{\citenamefont {Shor}(1995)}]{shor_scheme_1995}%
  \BibitemOpen
  \bibfield  {author} {\bibinfo {author} {\bibfnamefont {P.~W.}\ \bibnamefont {Shor}},\ }\bibfield  {title} {\bibinfo {title} {Scheme for reducing decoherence in quantum computer memory},\ }\href {https://doi.org/10.1103/PhysRevA.52.R2493} {\bibfield  {journal} {\bibinfo  {journal} {Phys. Rev. A}\ }\textbf {\bibinfo {volume} {52}},\ \bibinfo {pages} {R2493} (\bibinfo {year} {1995})}\BibitemShut {NoStop}%
\bibitem [{\citenamefont {Knill}(2005)}]{knill_quantum_2005}%
  \BibitemOpen
  \bibfield  {author} {\bibinfo {author} {\bibfnamefont {E.}~\bibnamefont {Knill}},\ }\bibfield  {title} {\bibinfo {title} {Quantum computing with realistically noisy devices},\ }\href {https://doi.org/10.1038/nature03350} {\bibfield  {journal} {\bibinfo  {journal} {Nature}\ }\textbf {\bibinfo {volume} {434}},\ \bibinfo {pages} {39} (\bibinfo {year} {2005})}\BibitemShut {NoStop}%
\bibitem [{\citenamefont {DiVincenzo}\ and\ \citenamefont {Aliferis}(2007)}]{divincenzo_effective_2007}%
  \BibitemOpen
  \bibfield  {author} {\bibinfo {author} {\bibfnamefont {D.~P.}\ \bibnamefont {DiVincenzo}}\ and\ \bibinfo {author} {\bibfnamefont {P.}~\bibnamefont {Aliferis}},\ }\bibfield  {title} {\bibinfo {title} {Effective {{Fault-Tolerant Quantum Computation}} with {{Slow Measurements}}},\ }\href {https://doi.org/10.1103/PhysRevLett.98.020501} {\bibfield  {journal} {\bibinfo  {journal} {Phys. Rev. Lett.}\ }\textbf {\bibinfo {volume} {98}},\ \bibinfo {pages} {020501} (\bibinfo {year} {2007})}\BibitemShut {NoStop}%
\bibitem [{\citenamefont {Landahl}\ \emph {et~al.}(2011)\citenamefont {Landahl}, \citenamefont {Anderson},\ and\ \citenamefont {Rice}}]{landahl_faulttolerant_2011}%
  \BibitemOpen
  \bibfield  {author} {\bibinfo {author} {\bibfnamefont {A.~J.}\ \bibnamefont {Landahl}}, \bibinfo {author} {\bibfnamefont {J.~T.}\ \bibnamefont {Anderson}},\ and\ \bibinfo {author} {\bibfnamefont {P.~R.}\ \bibnamefont {Rice}},\ }\bibfield  {title} {\bibinfo {title} {Fault-tolerant quantum computing with color codes},\ }\bibfield  {journal} {\bibinfo  {journal} {arXiv preprint}\ }\href {https://doi.org/10.48550/arXiv.1108.5738} {10.48550/arXiv.1108.5738} (\bibinfo {year} {2011}),\ \Eprint {https://arxiv.org/abs/1108.5738} {arXiv:1108.5738} \BibitemShut {NoStop}%
\bibitem [{\citenamefont {Fowler}\ \emph {et~al.}(2012)\citenamefont {Fowler}, \citenamefont {Mariantoni}, \citenamefont {Martinis},\ and\ \citenamefont {Cleland}}]{fowler_surface_2012}%
  \BibitemOpen
  \bibfield  {author} {\bibinfo {author} {\bibfnamefont {A.~G.}\ \bibnamefont {Fowler}}, \bibinfo {author} {\bibfnamefont {M.}~\bibnamefont {Mariantoni}}, \bibinfo {author} {\bibfnamefont {J.~M.}\ \bibnamefont {Martinis}},\ and\ \bibinfo {author} {\bibfnamefont {A.~N.}\ \bibnamefont {Cleland}},\ }\bibfield  {title} {\bibinfo {title} {Surface codes: {{Towards}} practical large-scale quantum computation},\ }\href {https://doi.org/10.1103/PhysRevA.86.032324} {\bibfield  {journal} {\bibinfo  {journal} {Phys. Rev. A}\ }\textbf {\bibinfo {volume} {86}},\ \bibinfo {pages} {032324} (\bibinfo {year} {2012})}\BibitemShut {NoStop}%
\bibitem [{\citenamefont {Devitt}\ \emph {et~al.}(2013)\citenamefont {Devitt}, \citenamefont {Munro},\ and\ \citenamefont {Nemoto}}]{devitt_quantum_2013}%
  \BibitemOpen
  \bibfield  {author} {\bibinfo {author} {\bibfnamefont {S.~J.}\ \bibnamefont {Devitt}}, \bibinfo {author} {\bibfnamefont {W.~J.}\ \bibnamefont {Munro}},\ and\ \bibinfo {author} {\bibfnamefont {K.}~\bibnamefont {Nemoto}},\ }\bibfield  {title} {\bibinfo {title} {Quantum error correction for beginners},\ }\href {https://doi.org/10.1088/0034-4885/76/7/076001} {\bibfield  {journal} {\bibinfo  {journal} {Rep. Prog. Phys.}\ }\textbf {\bibinfo {volume} {76}},\ \bibinfo {pages} {076001} (\bibinfo {year} {2013})}\BibitemShut {NoStop}%
\bibitem [{\citenamefont {Rist{\`e}}\ \emph {et~al.}(2020)\citenamefont {Rist{\`e}}, \citenamefont {Govia}, \citenamefont {Donovan}, \citenamefont {Fallek}, \citenamefont {Kalfus}, \citenamefont {Brink}, \citenamefont {Bronn},\ and\ \citenamefont {Ohki}}]{riste_realtime_2020}%
  \BibitemOpen
  \bibfield  {author} {\bibinfo {author} {\bibfnamefont {D.}~\bibnamefont {Rist{\`e}}}, \bibinfo {author} {\bibfnamefont {L.~C.~G.}\ \bibnamefont {Govia}}, \bibinfo {author} {\bibfnamefont {B.}~\bibnamefont {Donovan}}, \bibinfo {author} {\bibfnamefont {S.~D.}\ \bibnamefont {Fallek}}, \bibinfo {author} {\bibfnamefont {W.~D.}\ \bibnamefont {Kalfus}}, \bibinfo {author} {\bibfnamefont {M.}~\bibnamefont {Brink}}, \bibinfo {author} {\bibfnamefont {N.~T.}\ \bibnamefont {Bronn}},\ and\ \bibinfo {author} {\bibfnamefont {T.~A.}\ \bibnamefont {Ohki}},\ }\bibfield  {title} {\bibinfo {title} {Real-time processing of stabilizer measurements in a bit-flip code},\ }\href {https://doi.org/10.1038/s41534-020-00304-y} {\bibfield  {journal} {\bibinfo  {journal} {npj Quantum Inf}\ }\textbf {\bibinfo {volume} {6}},\ \bibinfo {pages} {71} (\bibinfo {year} {2020})}\BibitemShut {NoStop}%
\bibitem [{\citenamefont {Urbanek}\ \emph {et~al.}(2020)\citenamefont {Urbanek}, \citenamefont {Nachman},\ and\ \citenamefont {De~Jong}}]{urbanek_error_2020}%
  \BibitemOpen
  \bibfield  {author} {\bibinfo {author} {\bibfnamefont {M.}~\bibnamefont {Urbanek}}, \bibinfo {author} {\bibfnamefont {B.}~\bibnamefont {Nachman}},\ and\ \bibinfo {author} {\bibfnamefont {W.~A.}\ \bibnamefont {De~Jong}},\ }\bibfield  {title} {\bibinfo {title} {Error detection on quantum computers improving the accuracy of chemical calculations},\ }\href {https://doi.org/10.1103/PhysRevA.102.022427} {\bibfield  {journal} {\bibinfo  {journal} {Phys. Rev. A}\ }\textbf {\bibinfo {volume} {102}},\ \bibinfo {pages} {022427} (\bibinfo {year} {2020})}\BibitemShut {NoStop}%
\bibitem [{\citenamefont {Paetznick}\ and\ \citenamefont {Svore}(2014)}]{paetznick_repeatuntilsuccess_2014}%
  \BibitemOpen
  \bibfield  {author} {\bibinfo {author} {\bibfnamefont {A.}~\bibnamefont {Paetznick}}\ and\ \bibinfo {author} {\bibfnamefont {K.~M.}\ \bibnamefont {Svore}},\ }\bibfield  {title} {\bibinfo {title} {Repeat-{{Until-Success}}: {{Non-deterministic}} decomposition of single-qubit unitaries},\ }\href {https://doi.org/10.26421/QIC14.15-16-2} {\bibfield  {journal} {\bibinfo  {journal} {QIC}\ }\textbf {\bibinfo {volume} {14}},\ \bibinfo {pages} {1277} (\bibinfo {year} {2014})}\BibitemShut {NoStop}%
\bibitem [{\citenamefont {C{\'o}rcoles}\ \emph {et~al.}(2021)\citenamefont {C{\'o}rcoles}, \citenamefont {Takita}, \citenamefont {Inoue}, \citenamefont {Lekuch}, \citenamefont {Minev}, \citenamefont {Chow},\ and\ \citenamefont {Gambetta}}]{corcoles_exploiting_2021}%
  \BibitemOpen
  \bibfield  {author} {\bibinfo {author} {\bibfnamefont {A.~D.}\ \bibnamefont {C{\'o}rcoles}}, \bibinfo {author} {\bibfnamefont {M.}~\bibnamefont {Takita}}, \bibinfo {author} {\bibfnamefont {K.}~\bibnamefont {Inoue}}, \bibinfo {author} {\bibfnamefont {S.}~\bibnamefont {Lekuch}}, \bibinfo {author} {\bibfnamefont {Z.~K.}\ \bibnamefont {Minev}}, \bibinfo {author} {\bibfnamefont {J.~M.}\ \bibnamefont {Chow}},\ and\ \bibinfo {author} {\bibfnamefont {J.~M.}\ \bibnamefont {Gambetta}},\ }\bibfield  {title} {\bibinfo {title} {Exploiting {{Dynamic Quantum Circuits}} in a {{Quantum Algorithm}} with {{Superconducting Qubits}}},\ }\href {https://doi.org/10.1103/PhysRevLett.127.100501} {\bibfield  {journal} {\bibinfo  {journal} {Phys. Rev. Lett.}\ }\textbf {\bibinfo {volume} {127}},\ \bibinfo {pages} {100501} (\bibinfo {year} {2021})}\BibitemShut {NoStop}%
\bibitem [{\citenamefont {Pino}\ \emph {et~al.}(2021)\citenamefont {Pino}, \citenamefont {Dreiling}, \citenamefont {Figgatt}, \citenamefont {Gaebler}, \citenamefont {Moses}, \citenamefont {Allman}, \citenamefont {Baldwin}, \citenamefont {{Foss-Feig}}, \citenamefont {Hayes}, \citenamefont {Mayer}, \citenamefont {{Ryan-Anderson}},\ and\ \citenamefont {Neyenhuis}}]{pino_demonstration_2021}%
  \BibitemOpen
  \bibfield  {author} {\bibinfo {author} {\bibfnamefont {J.~M.}\ \bibnamefont {Pino}}, \bibinfo {author} {\bibfnamefont {J.~M.}\ \bibnamefont {Dreiling}}, \bibinfo {author} {\bibfnamefont {C.}~\bibnamefont {Figgatt}}, \bibinfo {author} {\bibfnamefont {J.~P.}\ \bibnamefont {Gaebler}}, \bibinfo {author} {\bibfnamefont {S.~A.}\ \bibnamefont {Moses}}, \bibinfo {author} {\bibfnamefont {M.~S.}\ \bibnamefont {Allman}}, \bibinfo {author} {\bibfnamefont {C.~H.}\ \bibnamefont {Baldwin}}, \bibinfo {author} {\bibfnamefont {M.}~\bibnamefont {{Foss-Feig}}}, \bibinfo {author} {\bibfnamefont {D.}~\bibnamefont {Hayes}}, \bibinfo {author} {\bibfnamefont {K.}~\bibnamefont {Mayer}}, \bibinfo {author} {\bibfnamefont {C.}~\bibnamefont {{Ryan-Anderson}}},\ and\ \bibinfo {author} {\bibfnamefont {B.}~\bibnamefont {Neyenhuis}},\ }\bibfield  {title} {\bibinfo {title} {Demonstration of the trapped-ion quantum {{CCD}} computer architecture},\ }\href {https://doi.org/10.1038/s41586-021-03318-4} {\bibfield  {journal} {\bibinfo  {journal} {Nature}\ }\textbf {\bibinfo {volume} {592}},\ \bibinfo {pages} {209} (\bibinfo {year} {2021})}\BibitemShut {NoStop}%
\bibitem [{\citenamefont {Yu}\ \emph {et~al.}(2025)\citenamefont {Yu}, \citenamefont {Yan}, \citenamefont {Biswas}, \citenamefont {Zhang}, \citenamefont {Harraz}, \citenamefont {Noel}, \citenamefont {Monroe},\ and\ \citenamefont {Kozhanov}}]{yu_insitu_2025}%
  \BibitemOpen
  \bibfield  {author} {\bibinfo {author} {\bibfnamefont {Y.}~\bibnamefont {Yu}}, \bibinfo {author} {\bibfnamefont {K.}~\bibnamefont {Yan}}, \bibinfo {author} {\bibfnamefont {D.}~\bibnamefont {Biswas}}, \bibinfo {author} {\bibfnamefont {V.~N.}\ \bibnamefont {Zhang}}, \bibinfo {author} {\bibfnamefont {B.}~\bibnamefont {Harraz}}, \bibinfo {author} {\bibfnamefont {C.}~\bibnamefont {Noel}}, \bibinfo {author} {\bibfnamefont {C.}~\bibnamefont {Monroe}},\ and\ \bibinfo {author} {\bibfnamefont {A.}~\bibnamefont {Kozhanov}},\ }\bibfield  {title} {\bibinfo {title} {In-situ mid-circuit qubit measurement and reset in a single-species trapped-ion quantum computing system},\ }\bibfield  {journal} {\bibinfo  {journal} {arXiv preprint}\ }\href {https://doi.org/10.48550/arXiv.2504.12544} {10.48550/arXiv.2504.12544} (\bibinfo {year} {2025}),\ \Eprint {https://arxiv.org/abs/2504.12544} {arXiv:2504.12544} \BibitemShut {NoStop}%
\bibitem [{\citenamefont {Carrera~Vazquez}\ \emph {et~al.}(2024)\citenamefont {Carrera~Vazquez}, \citenamefont {Tornow}, \citenamefont {Rist{\`e}}, \citenamefont {Woerner}, \citenamefont {Takita},\ and\ \citenamefont {Egger}}]{carreravazquez_combining_2024}%
  \BibitemOpen
  \bibfield  {author} {\bibinfo {author} {\bibfnamefont {A.}~\bibnamefont {Carrera~Vazquez}}, \bibinfo {author} {\bibfnamefont {C.}~\bibnamefont {Tornow}}, \bibinfo {author} {\bibfnamefont {D.}~\bibnamefont {Rist{\`e}}}, \bibinfo {author} {\bibfnamefont {S.}~\bibnamefont {Woerner}}, \bibinfo {author} {\bibfnamefont {M.}~\bibnamefont {Takita}},\ and\ \bibinfo {author} {\bibfnamefont {D.~J.}\ \bibnamefont {Egger}},\ }\bibfield  {title} {\bibinfo {title} {Combining quantum processors with real-time classical communication},\ }\href {https://doi.org/10.1038/s41586-024-08178-2} {\bibfield  {journal} {\bibinfo  {journal} {Nature}\ }\textbf {\bibinfo {volume} {636}},\ \bibinfo {pages} {75} (\bibinfo {year} {2024})}\BibitemShut {NoStop}%
\bibitem [{\citenamefont {Graham}\ \emph {et~al.}(2023)\citenamefont {Graham}, \citenamefont {Phuttitarn}, \citenamefont {Chinnarasu}, \citenamefont {Song}, \citenamefont {Poole}, \citenamefont {Jooya}, \citenamefont {Scott}, \citenamefont {Scott}, \citenamefont {Eichler},\ and\ \citenamefont {Saffman}}]{graham_midcircuit_2023}%
  \BibitemOpen
  \bibfield  {author} {\bibinfo {author} {\bibfnamefont {T.~M.}\ \bibnamefont {Graham}}, \bibinfo {author} {\bibfnamefont {L.}~\bibnamefont {Phuttitarn}}, \bibinfo {author} {\bibfnamefont {R.}~\bibnamefont {Chinnarasu}}, \bibinfo {author} {\bibfnamefont {Y.}~\bibnamefont {Song}}, \bibinfo {author} {\bibfnamefont {C.}~\bibnamefont {Poole}}, \bibinfo {author} {\bibfnamefont {K.}~\bibnamefont {Jooya}}, \bibinfo {author} {\bibfnamefont {J.}~\bibnamefont {Scott}}, \bibinfo {author} {\bibfnamefont {A.}~\bibnamefont {Scott}}, \bibinfo {author} {\bibfnamefont {P.}~\bibnamefont {Eichler}},\ and\ \bibinfo {author} {\bibfnamefont {M.}~\bibnamefont {Saffman}},\ }\bibfield  {title} {\bibinfo {title} {Midcircuit {{Measurements}} on a {{Single-Species Neutral Alkali Atom Quantum Processor}}},\ }\href {https://doi.org/10.1103/PhysRevX.13.041051} {\bibfield  {journal} {\bibinfo  {journal} {Phys. Rev. X}\ }\textbf {\bibinfo {volume} {13}},\ \bibinfo {pages} {041051} (\bibinfo {year} {2023})}\BibitemShut {NoStop}%
\bibitem [{git()}]{github_parallel_hadamard_tests}%
  \BibitemOpen
  \href@noop {} {}\bibinfo {note} {{GitHub, \url{https://github.com/soichiro524/parallel-hadamard-tests}}}\BibitemShut {NoStop}%
\bibitem [{ibm()}]{ibm_quantum_workload}%
  \BibitemOpen
  \href@noop {} {}\bibinfo {note} {{See \url{https://quantum.cloud.ibm.com/docs/en/guides/estimate-job-run-time}}}\BibitemShut {NoStop}%
\bibitem [{aws()}]{aws_braket_pricing}%
  \BibitemOpen
  \href@noop {} {}\bibinfo {note} {{See \url{https://aws.amazon.com/braket/pricing/}}}\BibitemShut {NoStop}%
\bibitem [{qbr()}]{qbraid_pricing}%
  \BibitemOpen
  \href@noop {} {}\bibinfo {note} {{See \url{https://docs.qbraid.com/home/pricing}}}\BibitemShut {NoStop}%
\end{thebibliography}%

\end{document}